\documentclass[iop,onecolumn]{emulateapj}
\shorttitle{Coevolution of Black Holes and Massive Galaxies at High $z$}
\shortauthors{A. Lapi et al.}

\begin{document}
\title{The Coevolution of Supermassive Black Holes and Massive Galaxies
at High Redshift}
\author{A. Lapi\altaffilmark{1,2}, S. Raimundo\altaffilmark{1}, R.
Aversa\altaffilmark{1},
Z.-Y. Cai\altaffilmark{1,3}, M. Negrello\altaffilmark{4}, A.
Celotti\altaffilmark{1,5,6}, G. De Zotti\altaffilmark{1,4}, L.
Danese\altaffilmark{1,6}} \altaffiltext{1}{SISSA, Via Bonomea 265, 34136
Trieste, Italy} \altaffiltext{2}{Dip. Fisica, Univ. `Tor Vergata', Via
Ricerca Scientifica 1, 00133 Roma, Italy} \altaffiltext{3}{Dept. of Astron.
and Inst. of Theor. Physics and Astrophysics, Xiamen Univ., 361005 Xiamen,
China} \altaffiltext{4}{INAF-Osservatorio Astronomico di Padova, Vicolo
dell'Osservatorio 5, 35122 Padova, Italy}\altaffiltext{5} {INAF-Osservatorio
Astronomico di Brera, via E. Bianchi 46, 23807 Merate, Italy}
\altaffiltext{6}{INFN-Sezione di Trieste, via Valerio 2, 34127, Trieste,
Italy}

\begin{abstract}
We exploit the recent, wide samples of far-infrared (FIR) selected galaxies
followed-up in X rays and of X-ray/optically selected active galactic
nuclei (AGNs) followed-up in the FIR band, along with the classic data on
AGN and stellar luminosity functions at high redshift $z\gtrsim 1.5$, to
probe different stages in the coevolution of supermassive black holes (BHs)
and host galaxies. The results of our analysis indicate the following
scenario: (i) the star formation in the host galaxy proceeds within a
heavily dust-enshrouded medium at an almost constant rate over a timescale
$\lesssim 0.5-1$ Gyr, and then abruptly declines due to quasar feedback;
over the same timescale, (ii) part of the interstellar medium loses angular
momentum, reaches the circum-nuclear regions at a rate proportional to the
star formation and is temporarily stored into a massive
reservoir/proto-torus wherefrom it can be promptly accreted; (iii) the BH
grows by accretion in a self-regulated regime with radiative power that can
slightly exceed the Eddington limit $L/L_{\rm Edd}\lesssim 4$, particularly
at the highest redshifts; (iv) for massive BHs the ensuing energy feedback
at its maximum exceeds the stellar one and removes the interstellar gas,
thus stopping the star formation and the fueling of the reservoir; (v)
afterwards, if the latter has retained enough gas, a phase of
supply-limited accretion follows exponentially declining with a timescale
of about $2$ $e$-folding times. We also discuss how the detailed properties
and the specific evolution of the reservoir can be investigated via
coordinated, high-resolution observations of starforming, strongly-lensed
galaxies in the (sub-)mm band with \textsl{ALMA} and in the X-ray band with
\textsl{Chandra} and the next generation X-ray instruments.
\end{abstract}


\section{Introduction}
The joint formation and evolution of galaxies and active galactic
nuclei/quasars (AGNs/QSOs) is still a major problem in astrophysics and
cosmology. One of the key points has been the discovery, via stellar/gas
dynamics and photometric observations, that local elliptical-type galaxies
(ETGs) and massive bulges exhibit at their centers a massive dark object
(MDO) endowed with a mass $M_{\bullet}\sim 10^{-3}\,M_{\star}$ proportional
to the mass in \emph{old} stars or to the $K$-band luminosity (Kormendy \&
Richstone 1995; Magorrian et al. 1998; Gebhardt et al. 2000a; Marconi \& Hunt
2003; H\"{a}ring \& Rix 2004; McLure \& Dunlop 2004; Ferrarese \& Ford 2005;
Graham 2007; Sani et al. 2011; Beifiori et al. 2012; McConnell \& Ma 2013;
Kormendy \& Ho 2013).

Indeed, the hypothesis that MDOs are the black hole (BH) remnants of a past
QSO activity was one of the major motivation for their searches, following
the suggestion that QSOs are powered by gas accretion onto supermassive BHs
(Salpeter 1964; Zel'dovich \& Novikov 1964) and the integral argument
presented by Soltan (1982). Salucci et al. (1999) demonstrated that the mass
function of MDOs in local ETGs and bulges, derived on the basis of the local
$M_{\bullet}-M_{\star}$ correlation, is very close to the mass function of
the accreted matter estimated by exploiting the redshift-dependent AGN/QSO
luminosity function, if the paradigm of accretion disc onto a BH (with
reasonable matter-radiation conversion efficiency and limiting luminosity) is
assumed. Moreover, from the analysis of the orbits for a number of individual
stars in the central region of our Galaxy the presence of a supermassive BH
with mass $M_{\rm BH}\approx 4\times 10^6\, M_{\odot}$ has been established
at a very high confidence level (Genzel et al. 2010). Once MDOs are
identified as BHs, the $M_{\bullet}-M_{\star}$ relation translates
into a relationship between the BH mass $M_{\rm BH}$ and the stellar mass
$M_{\star}$, suggesting that the star formation on large (i.e., kpc) scale in
(proto-)ETGs and the accretion onto the BHs on much smaller (sub-pc) scale
must in a way talk to each other (e.g., Alexander \& Hickox 2012).

The luminosity/mass of the stellar component is not the only global property
of the local ETGs that correlates with the central BH mass. In fact, the
relation between the mass of the central BH and the stellar velocity
dispersion $\sigma_\star$ or, more recently, the velocity dispersion of the
globular cluster system, has been studied and it appears to be somewhat less
dispersed than that with the stellar luminosity/mass (Ferrarese \& Merritt
2000; Gebhardt et al. 2000b; Tremaine et al. 2002; G\"{u}ltekin et al. 2009;
McConnell et al. 2011; Graham et al. 2011; Beifiori et al. 2012; Sadoun \&
Colin 2012; McConnell \& Ma 2013; Pota et al. 2013; Kormendy \& Ho 2013).
This relationship is possibly imprinted by the QSO feedback (but also stellar
feedback can contribute); such an idea has been proposed by Silk \& Rees
(1998) based on an energy argument and by Fabian (1999) based on a momentum
one, and further developed by King (2003, 2005) and by Murray et al. (2005).

Also correlations between the central BH mass and the light profiles of the
host galaxies have been explored (Graham et al. 2001; Lauer et al. 2007;
Graham \& Driver 2007; Kormendy \& Bender 2009); these mirror the complexity
of the phenomena related to the coupled evolution of the host galaxy and its
central active nucleus. Comprehensive discussions of the subject have been
presented by Beifiori et al. (2012) and Kormendy \& Ho (2013).

The high degree of correlation, though with significant dispersion, between
the local values of $M_{\rm BH}$ and $M_{\star}$, or of $M_{\rm BH}$ and
$\sigma_\star$, strongly suggests that the star formation on galactic scales,
the accreted mass onto the central BH and the ensuing radiated power from the
AGN have to be in a way coordinated. However, this is a reductive conclusion,
since star formation, gas accretion and radiative feedback are quite complex
physical processes occurring on very different spatial and temporal scales.
Each of these phenomena is \textit{per se} a challenging problem to treat.

While some hints on the BH/galaxy coevolution can be derived by focusing onto
their low-$z$ properties (e.g., Kormendy \& Ho 2013), in this paper we aim at
showing that the wide samples of AGNs and host galaxies observed at high
redshift ($z\gtrsim 1.5$) already yield important information on how and when
the formation of the oldest stellar population in ETGs/bulges and the BH
growth influence each other. As a matter of fact, both the nuclear activity
(Granato et al. 2001; Richards et al. 2006; Driver et al. 2013; Kulier et al.
2013) and the formation of the old stellar populations in ETGs peak at
$z\approx 2$ (e.g., Fardal et al. 2007; Driver et al. 2013).

We leave aside the second stage of star formation in galactic \emph{thin}
discs (see Cook et al. 2009; Driver et al. 2013) as well as the low redshift
fading activity of AGNs, since there is practically no correlation between
the disc luminosity/mass and the BH mass (e.g., Kormendy \& Bender 2013;
Kormendy \& Ho 2013). Correspondingly, low redshift AGNs exhibit in general
small Eddington ratios (e.g., Vestergaard 2004; Vestergaard \& Osmer 2009;
Kelly \& Shen 2013), with the possible exception of the Narrow Line Seyfert 1
galaxies. A likely interpretation of this fading AGN phase is that it is
mainly due to short episodes of accretion onto a pre-existing supermassive BH
(Cavaliere \& Vittorini 2000; Ballo et al. 2007; Simmons et al. 2011; Shankar
2009; Shankar et al. 2013).

From a statistical point of view, the coevolution of the host galaxies and
their supermassive BHs could be reconstructed if the stellar mass/star
formation rate (SFR) distribution for the hosts and the BH mass/accretion
rate distribution for the AGNs/QSOs at different redshifts were available.
These pieces of information would provide insights at least on space- and
time-averaged quantities. Progress in this respect has been tremendous in the
last decade and nowadays we have sound estimates even at substantial redshift
of:

\begin{itemize}

\item luminosity/stellar mass functions (e.g., Stark et al. 2009;
    Marchesini et al. 2009, 2010; Cirasuolo et al. 2010; Ilbert et al.
2013);

\item far infrared (FIR) luminosity/SFR function in massive galaxies (e.g.,
    Eales et al. 2010; Gruppioni et al. 2010, 2013; Lapi et al. 2011);

\item BH masses (see Shen 2013 for a comprehensive review) and Eddington
    ratios (see Kelly \& Shen 2013 and references therein);

\item luminosity functions of AGNs/QSOs in the X-ray (e.g., Ueda et al.
    2003; Aird et al. 2008, 2010; Fiore et al. 2012a) and in the optical
    bands (e.g., Pei 1995; Wolf et al. 2003; Hunt et al. 2004; Richards et
    al. 2005, 2006; Fan et al. 2006; Fontanot et al. 2007; Croom et al.
    2009; Jiang et al. 2009; Willott et al. 2010; Masters et al. 2012; Ross
    et al. 2013).

\end{itemize}

On top of that, recently the coevolution has been also explored by searching
for nuclear activity in starforming galaxies and, viceversa, searching for
star formation in AGNs/QSOs. More in detail, large statistics have been
recently obtained on:

\begin{itemize}

\item nuclear activity, by exploiting the follow-up in X rays of galaxies
    with large SFR mainly selected at FIR/sub-mm wavelengths or in the
    $K-$band (e.g., Borys et al. 2005; Alexander et al. 2005, 2008; Laird
    et al. 2010; Symeonidis et al. 2010; Xue et al. 2010; Georgantopoulos
    et al. 2011; Carrera et al. 2011; Melbourne et al. 2011; Rafferty et
    al. 2011; Mullaney et al. 2012a; Johnson et al. 2013; Wang et al. 2013)
    and of galaxies starforming at a lower rate hence more easily selected
    at UV wavelengths via the stacking technique (e.g., Fiore et al. 2012b;
    Treister et al. 2011; Willott 2011; Basu-Zych et al. 2013);

\item the star formation in AGN host galaxies, by exploiting the follow-up
    at FIR and (sub-)mm wavelengths of X-ray selected AGNs (e.g., Page et
    al. 2004, 2012; Stevens et al. 2005; Lutz et al. 2010; Shao et al.
    2010; Mainieri et al. 2011; Harrison et al. 2012; Mullaney et al.
    2012b; Rosario et al. 2012; Rovilos et al. 2012; Santini et al. 2012)
    and of optically selected QSOs (Omont et al. 1996, 2001, 2003; Carilli
    et al. 2001; Priddey et al. 2003; Wang et al. 2008; Walter et al. 2009;
    Serjeant et al. 2010; Bonfield et al. 2011; Mor et al. 2012).

\end{itemize}

These two complementary blocks of observations are of paramount relevance in
determining the way stellar and BH mass grew at early times in the ETG
progenitors. In the present paper we will exploit a basic model to provide a
definite descriptions of the SFR and AGN lightcurves in terms of a few
physical parameters. We will show that the comparison with the current data
can constrain the model, and thus clarify the main aspects of the galaxy/AGN
coevolution process. We will also point out that additional observations in
X-ray, optical and FIR bands are strongly required in order to test the
overall picture in detail.

The paper is organized as follows. In \S~2 we present a simple prescription
for the time-averaged evolution of the SFR and BH accretion rate, following
the guidance of a basic physical model, and show that it fits the
statistics of the SFR (i.e., the FIR luminosity function) and the stellar 
mass
function at high redshift. \S~3 is devoted to show how powerful are the data
obtained by X-ray follow-up of FIR selected starforming galaxies and by FIR
follow-up of X-ray and optically selected AGNs/QSOs in constraining the
model. The comparison of our results with the AGN/QSO luminosity functions in
the X-ray and optical bands is also presented. In \S~4 we show that
additional observations, somewhat less systematic but of great relevance such
as the BH to stellar mass ratio in optically selected QSOs, the
relative abundance of obscured to unobscured AGNs/QSOs, and the measured QSO
outflow rates, compare well with the predictions of the model. In \S~5 we
discuss the prospects for direct detection of the large gas reservoir around
the supermassive BHs predicted by the model. We also discuss our
understanding of the gas path from the interstellar medium to the accretion
disc around the supermassive BHs. In \S~6 we summarize our findings.

Throughout this work we adopt the concordance cosmology (see \textsl{Planck}
Collaboration 2013), i.e., a flat universe with matter density parameter
$\Omega_M=0.31$, Hubble constant $H_0=100\, h$ km s$^{-1}$ Mpc$^{-1}$ with
$h=0.67$, and mass variance $\sigma_8=0.82$ on a scale of $8\, h^{-1}$ Mpc.
Stellar masses and FIR luminosities (or conversely SFRs) of galaxies are
evaluated assuming the Chabrier's (2003) initial mass function (IMF) and the
spectral energy distribution (SED) of SMM J2135-0102, a typical high-redshift
starbursting galaxy (very similar to the local ULIRG Arp220; see Lapi et al.
2011 for details). Specifically, the luminosity associated to the SFR reads
$3.2\times 10^{43}\, (\dot M_\star/M_\odot\,\mathrm{yr}^{-1})$ erg s$^{-1}$,
which practically coincides with the FIR luminosity $L_{\rm FIR}$ if dust
absorbs almost all of the stellar emission. By FIR we consider the restframe
wavelength range from $\lambda\approx 40\,\mu$m to $\lambda\approx 500\,
\mu$m. In general, we can assume that this spectral region is dominated by
the emission of dust associated with star formation, although a contribution
from a torus around an AGN may be not completely negligible (e.g., Granato \&
Danese 1994; Leipski et al. 2013). As to the nuclear emission, we will
indicate with $L_{\rm AGN}$ the bolometric output, and with $L_B=L_{\rm
AGN}/k_B$, $L_X=L_{\rm AGN}/k_X$ the powers in the optical and in the $2-10$
keV X-ray bands, adopting the bolometric corrections $k_B$ and $k_X$ by
Hopkins et al. (2007; see also Marconi et al. 2004, Vasudevan \& Fabian
2007). The values depend on the AGN luminosity (or on the Eddington ratio),
and span the ranges $k_B\sim 8-15$ and $k_X\sim 15-100$ for $L_{\rm AGN}\sim
10^{43}-10^{47}$ erg s$^{-1}$. Note that we will restrict ourselves to
$L_X\gtrsim 10^{42}$ erg s$^{-1}$, to avoid appreciable contaminations of the
X-ray emission by star formation (see Symeonidis et al. 2011).

\section{A basic model}

The observations mentioned in \S~1 can be exploited to cast light on how the
relationships between star formation and BH accretion in primeval galaxies
have been set up. To this purpose we need to exploit a simple but physically
motivated description of the SFR and of the accretion rate onto the BH as a
function of galactic time within dark matter halos of given mass and
formation redshift. Then the SFR is converted into FIR luminosity, under the
assumption that most of the star formation in massive galaxies occurs in a
dusty environment. The accretion rate is converted into bolometric luminosity
of the AGN by assuming a matter-to-radiation conversion efficiency, and
eventually into luminosity at specific wavelengths through bolometric
corrections. The light curves of the host galaxy and of the AGN are then
exploited for the computation of various statistics to be compared with the
data.

We take as a guidance the framework originally proposed in Granato et al.
(2004), which has been successful in reproducing the statistics of galaxies
selected at $850\,\mu$m, of passively evolving galaxies, and of AGNs/QSOs at
substantial redshift. Lapi et al. (2006, 2011) and Cai et al. (2013) have
further developed the original formulation of the model, and showed that it
fits the galaxy number counts at (sub-)mm wavelengths (e.g., Clements et al.
2010; Vieira et al. 2010, 2013) and the luminosity functions of sub-mm
selected galaxies (Eales et al. 2010; Lapi et al. 2011; Gruppioni et al.
2013). Furthermore, the model has been used to estimate the number of sub-mm
selected gravitationally lensed galaxies (Perrotta et al. 2003; Negrello et
al. 2007; Lapi et al. 2012), a prediction fully confirmed by observations
(Negrello et al. 2010; Gonzalez-Nuevo et al. 2012; Weiss et al. 2013). Even
the correlation function of both QSOs and sub-mm selected galaxies are very
well reproduced, ensuring that the model correctly locate galaxies and QSOs
into massive dark matter halos (Xia et al. 2012).

The model is based on a few assumptions, that we recall next. Concerning the
formation of the dark matter hosts, it exploits the outcomes of many
intensive $N-$body simulations and semianalytic studies (e.g., Zhao et al.
2003; Wang et al. 2011; Lapi \& Cavaliere 2011). These have recognized that
massive halos undergo an early phase of \emph{fast} collapse, during which
the central regions reach rapidly a dynamical quasi-equilibrium; a subsequent
slower accretion phase mainly affects the halo outskirts. The
transition between the two phases can be identified with the \emph{formation}
redshift. The halo formation rates at redshift $z\gtrsim 1.5$ and for halo
masses $M_{\rm H}\gtrsim 10^{10}\, M_{\odot}$ derived from cosmological
$N-$body simulations are very well approximated by the analytical formulae
given in Lapi et al. (2013), which have been exploited in the present work.

The mass of baryons associated to a dark halo is about $20\%$ of the total
one. It has been shown that in the local Universe the largest ratio of
stellar-to-halo mass amounts to $M_{\star}/M_{\rm H}\lesssim 0.03$ in
galactic halos of mass $M_{\rm H}\approx 10^{12}\, M_{\odot}$ (e.g., Vale \&
Ostriker 2004; Shankar et al. 2006; Guo et al. 2010; Moster et al. 2010,
2013). Only $\lesssim 15\%$ of the available baryons are finally locked up
into stars and such a percentage can be $\lesssim 10\%$ at higher redshift
(Moster et al. 2013). Assuming a Chabrier (2003) IMF the fraction of baryons
passed through the stellar cycle at high redshift is $\lesssim 20\%$. Thus
very likely $\lesssim 40\%$ of the baryons associated to dark halos are
possibly involved in the formation of galaxies. This corresponds to the mass
which is involved in the early fast collapse. It is worth noticing that the
average inefficiency of star formation is highlighted by the small ratio of
density parameters in stars and baryons $\Omega_{\star}/\Omega_{b}\approx
5\%$ (e.g., Fukugita \& Peebles 2004; Shankar et al. 2006; Li \& White 2009),
corresponding to $\Omega_{\star}/\Omega_{\rm DM}\approx 1\%$ (e.g., Moustakas
et al. 2013).

The fast collapse is expected to be inhomogeneous and to proceed through
generation of clumps (see Ceverino et al. 2010; Bournaud et al. 2014).
Gravitational instabilities, tidal torques and dynamical friction among the
starforming clumps, coupled with large gas flows generated by stellar winds
and supernova explosions, strongly favor the dissipation of the angular
momentum left over by the dynamical relaxation occurred during the fast
collapse (see Noguchi 1999; Immeli et al. 2004; Elmegreen et al. 2008). In
this framework, our model treats the coevolution of the galaxy and its
central BH by assuming that the angular momentum of the baryonic clumps is
dissipated by such mechanisms on a timescale comparable with or shorter than
the duration $\Delta t_{\rm burst}$ of the starbursting activity. Therefore
the evolution of massive galaxies is dictated by secular internal processes
occurring in the central regions of galactic halos. Since these processes
start, major mergers are no longer relevant for the galaxy/QSO evolution and
are neglected. This is a crucial feature of our model, as opposed to other
treatments (for a list, see Scannapieco et al. 2012; for a critical view, see
Frenk \& White 2012 and Kaviraj et al. 2013).

The galaxy is considered to be constituted by two gas phases, the infalling
hot/warm gas with mass $M_{\rm inf}$ and the cold gas with mass $M_{\rm
cold}$, and by stars (forming and dying). The hot/warm gas condenses into the
cold one at a rate $\dot{M}_{\rm cold}=M_{\rm inf}/t_{\rm cond}$, where the
timescale $t_{\rm cond}$ is the longer between the free fall and the cooling
timescales, computed at the virial radius by assuming for the hot/warm gas a
clumping factor $\approx 7$ (see Lapi et al. 2006); an analytical
approximation reads
\begin{equation}
t_{\rm cond}\approx 7\times 10^8\left(1+z\over 3.5\right)^{-1.5}\, \left
(M_{\rm H}\over 10^{12}\ M_{\odot}\right)^{0.2}~\mathrm{yr}~.
\end{equation}

The average SFR is computed as $\dot{M}_{\star}=M_{\rm cold}/t_{\rm
cond}\times s$, and is assumed to occur on a timescale shorter than the
condensation time since the cold gas is expected to fragment and to increase
its clumping factor well above that of the hot/warm component. The factor $s$
is adjusted at a value around $5$ to fit the aforementioned wide sets of
data, and in particular the sub-mm luminosity functions and counts.

On the other hand, the feedback from stars, proportional to the number of
supernovae (SN) and hence to the SFR, removes the cold gas at a rate
\begin{equation}
\dot{M}_{\rm cold}^{\rm SN}={N_{\rm SN}\,\epsilon_{\rm SN}\, E_{\rm SN}\over
E_{\rm bind}}\, \dot{M}_{\star}~,
\end{equation}
where $N_{\rm SN}\approx 1.4\times 10^{-2}/M_\odot$ is the number of SNae per
unit solar mass condensed into stars, $E_{\rm SN}\approx 10^{51}$ erg is the
kinetic energy released per SN explosion, $E_{\rm bind}\approx 3\times
10^{14}$ $(M_{\rm H}/10^{12}\,M_\odot)^{2/3}$ $[(1+z)/3.5]$ cm$^2$ s$^{-2}$
is the specific binding energy of the gas in the halo; $\epsilon_{\rm
SN}=0.05$ is the standard value adopted for the SN feedback efficiency (e.g.,
White \& Frenk 1991; Cole et al. 2000).

Granato et al. (2004) were among the first to work out a quantitative model
for the feedback from nuclear activity. According to this model, the feedback
is able to affect the most external parts of the galaxy and to act on the
hot/warm and cold gas in a way proportional to their fractional mass. For the
cold gas the rate of removal is
\begin{equation}
\dot{M}_{\rm cold}^{\rm QSO}=f_{\rm cold}\frac {L_{\rm QSO}}{E_{\rm bind}}~,
\end{equation}
where $f_{\rm cold}$ is the cold to total gas mass fraction, and $L_{\rm
QSO}\approx 2\times 10^{44} \, \epsilon _{\rm QSO}\, \left(M_{\rm BH}/10^8 \,
M_{\odot}\right)^{3/2}$ erg s$^{-1}$ is the mechanical energy delivered from
the QSO in terms of a strength parameter $\epsilon_{QSO}$; values
$\epsilon_{QSO}\sim 1$ are required to reproduce the bright end of the
stellar mass function and of the QSO luminosity functions (see Lapi et al.
2006 and Cai et al. 2013 for details). A similar expression holds for the
feedback on the hot/warm component.

Examples of the evolution with galactic age of the FIR luminosity associated
to the SFR are illustrated in Fig.~1. The curves refer to halo masses $M_{\rm
H}=2$ and $6\times 10^{12}\, M_{\odot}$ at $z=2$. For these massive galaxies,
the SFR can be approximated for most of the time ($\gtrsim 80\%$ of the burst
duration) by a constant with a dependence on halo mass and redshift given by
\begin{equation}
\dot{M}_\star = 70\, \left({M_{\rm H}\over 10^{12}\,
M_{\odot}}\right)\,\left({1+z\over 3.5}\right)^{2.1}\,
M_{\odot}~\mathrm{yr}^{-1}.
\end{equation}
The duration of the main burst is approximately $\Delta t_{\rm burst}\approx
7\times 10^{8}\, [(1+z)/3.5]^{-1.5}$ yr.

In the model the approximate constancy of the SFR mirrors that of the mass in
cold gas. In turn, the latter stems from the balance among: (i) the
condensation rate of warm/hot gas into the cold phase; (ii) the rate of star
formation; (iii) the rate of mass restitution from stellar evolution to the
cold gas phase; (iv) the rate of cold gas removal by stellar feedback (SN and
stellar winds). We stress that such a balance also ensures a rapid increase
of the metallicity in the cold gas and hence in the formed stars. A rapid
chemical enrichment is fundamental in creating conditions apt to a rapid
formation of large amounts of dust in massive galaxies at high redshift; in
combination with the relatively short duration $\lesssim 0.5-1$ Gyr of the
star formation the enrichment is also relevant to produce chemical abundances
and $\alpha-$enhancement in agreement with those observed for local massive
ETGs (see Silva et al. 1998; Granato et al. 2004; Lapi et al. 2006, 2011).

The FIR lightcurves depend on halo mass and redshift; to derive the
luminosity function of starforming galaxies in the FIR the galaxy luminosity
must be convolved with the halo formation rate (see Appendix). The outcome is
illustrated in Fig.~2, which shows that the model nicely reproduces the
observed luminosity functions of starforming galaxies at high redshift
$z\gtrsim 1.5$. We stress that also galaxy counts, clustering properties and
the cosmic infrared background autocorrelation function at (sub-)mm
wavelengths can be reproduced by exploiting a quite standard spectral energy
distribution for dust emission, similar to that of local ULIRGs (see Lapi et
al. 2011; Cai et al. 2013).

The stellar mass is linearly increasing with time for $t\lesssim \Delta
t_{\rm burst}$ and can be approximated by
\begin{equation}
M_{\star}(t)\approx 5\times 10^{10}\, \left({t\over \Delta t_{\rm
burst}}\right)\, \left({M_{\rm H}\over 10^{12}\, M_{\odot}}\right)\,
\left({1+z\over 3.5}\right)^{0.6}\, M_{\odot}~
\end{equation}
if a Chabrier (2003) mass function is assumed; due to gas recycling, the
present-day mass in stars is about half of the mass in stars formed before
$\Delta t_{\rm burst}$. The evolution of the stellar mass is presented in
Fig.~3, while the stellar mass function at the relevant redshifts yielded by
the model is confronted with observations in Fig.~4. The agreement
corroborates the physical assumptions of the model. Its effectiveness resides
in the capability of locating the quite large observed SFR at high redshift
in the most massive, strongly clustered halos.

As for the nuclear activity, the gas flowing toward the BH through the
accretion disc produces a \emph{time-averaged} radiative output
\begin{equation}
\epsilon\, c^2\, \dot{M}_{\rm accr}=L_{\rm AGN}=\lambda\, L_{\rm Edd}~,
\end{equation}
where $\dot M_{\rm accr}$ is the accretion rate,
$0.06\lesssim\epsilon\lesssim 0.3$ is the mass to radiation conversion
efficiency ($\epsilon=0.15$ has been assumed in this work, see the discussion
in \S~5.2.3), and $\lambda\, L_{\rm Edd}$ is the luminosity in terms of the
Eddington one. Lapi et al. (2006) have shown that the Eddington ratio has on
average to increase with redshift (for $z\ga 1.5$) as
$\lambda(z)=-1.15+0.75\,(1+z)$ in order to reproduce the QSO luminosity
functions for $1.5\lesssim z \lesssim 6$. The maximum allowed value
$\lambda\lesssim 4$ is in good agreement with recent estimates from
observations (see Kelly \& Shen 2013). If $\epsilon$ and $\lambda$ are
assumed to stay on average constant during most of the accretion time (we
defer to \S~5.2.3 the discussion of these assumptions), Eq.~(6) implies that
the BH mass and luminosity grow exponentially on a timescale $\tau_{\rm
ef}=\epsilon/\lambda (1-\epsilon)\times t_{\rm Edd}$ with $t_{\rm Edd}\approx
4.5\times 10^8$ yr; we recall that $\dot{M}_{\rm BH}=(1-\epsilon)\,
\dot{M}_{\rm accr}$.

The mean observed ratio between the stellar and BH mass in local massive ETGs
and bulges is $\Gamma_0= M_{\rm BH}/M_{\star}\approx 3\times 10^{-3}$ though
with a large scatter of $\approx 0.4$ dex (e.g., Magorrian et al. 1998;
Gebhardt et al. 2000a; Marconi \& Hunt 2003; H\"{a}ring \& Rix 2004;
Ferrarese \& Ford 2005; Bennert et al. 2011; Graham et al. 2011; Beifiori et
al. 2012; McConnell \& Ma 2013; Kormendy \& Ho 2013). This suggests that the
cold gas inflows from galactic scales to the central regions ($\lesssim 100$
pc) at a rate proportional to the SFR, i.e.,
\begin{equation}
\dot{M}_{\rm inflow}=\alpha_{\rm res}\times 10^{-3}\,\dot{M}_{\star}~.
\end{equation}
The efficiency $\alpha_{\rm res}(M_{\rm H},z)$ is possibly a function of halo
mass and formation redshift, and depends on the particular process considered
for the gas to lose its angular momentum on large scales (see Granato et al.
2004; Lapi et al. 2006). The above proportionality is also supported, though
with a substantial scatter, by several numerical simulations as recalled in
\S~5.2. In particular, Hopkins \& Quataert (2010, 2011) show that
gravitational instabilities can trigger star formation and at the same time
funnel part of the interstellar medium (ISM) toward the central region
($\lesssim 100$ pc), where the gas attains large surface densities.

At this stage the gas inflowing from the ISM can be either directly accreted
onto the BH or can pile up into a \emph{reservoir}. As we shall see in \S~3,
the data do not support a time-averaged proportionality between the
bolometric AGN luminosity and the FIR emission associated with the SFR of the
host galaxy.

In addition, if one assume a conceivably small seed BH mass $M_{\rm BH}\sim
10^2-10^3\, M_{\odot}$ (for a review, see Volonteri 2010), the large and
constant SFR implies the rate
\begin{equation}
\dot{M}_{\rm res}=\dot{M}_{\rm inflow}-\dot{M}_{\rm accr} = \alpha_{\rm
res}\,\times 10^{-3}\,\dot{M}_{\star}-{\lambda\over \epsilon}\,
{L_{\rm Edd}\over c^2}~,
\end{equation}
to be positive for many $e-$folding times, allowing a reservoir to form and
its mass to significantly increase; e.g., for a $\dot M_\star\ga 10^2\,
M_\odot$ yr$^{-1}$ and a seed BH mass $M_{\bullet}\approx 10^3\, M_\odot$,
one has the formation of a reservoir ($\dot M_{\rm res}\ga 0$) unless
$\lambda/\epsilon\ga 10^5$ is exceptionally high (cf. discussion in
\S~5.2.3). Examples of the time evolution for the reservoir and BH masses are
plotted in Fig.~3. The reservoir can be related to the so called torus,
observed in nearby AGNs and called for to explain the AGN phenomenology
(e.g., Antonucci 1993; Granato \& Danese 1994; Urry \& Padovani 1995).
Observations, at least of low redshift objects (see, e.g., Davies et al.
2007; Muller-Sanchez et al. 2009; Krips et al. 2011; Diamond-Stanic \& Rieke
2012; Sani et al. 2012; Storchi-Bregmann et al. 2012; H\"{o}nig et al. 2013),
have revealed the presence of such reservoirs, with sizes from a few to
several tens of parsecs; the reservoirs have been found to be rich in
molecular gas and dust, and often accompanied by localized star formation.

Additional physical mechanisms, such as turbulence generated by SN explosion,
gravitational tidal torques, dynamical friction among clumps, can further
reduce the angular momentum of the gas making it ready to flow through the
accretion disc (see \S~5.2 for a discussion). The gas piled up in the
reservoir is large enough to sustain accretion during the two final
$e$-folding times, when the QSO reaches its maximum luminosity and the BH
acquires most of its final mass. During this short epoch lasting $\lesssim
10^8$ yr powerful winds can be generated, capable of removing most of the gas
within the host galaxy, of quenching the star formation, and thus stopping
the fueling of the reservoir. The gas remaining in the reservoir still flows
toward the BH, though at a decreasing rate; thus the accretion luminosity
decreases to only a fraction of the Eddington limit.

If, and only if, a significant fraction of the gas stored in the reservoir is
finally accreted onto the seed BH, then the correlation between the mass in
old stars and the BH mass found in local ETGs is automatically set up. For a
Chabrier IMF only about half (in detail, a factor $1/1.7$) of the mass in
stars formed early is presently still there; this implies that $M_{\rm
BH}\approx 1.7\times \alpha_{\rm res}\times 10^{-3}\, M_{\star}$ holds, so
average values $\alpha_{\rm res}\sim 2$ would yield the local ratio
$\Gamma_0$. By the same token, $\alpha_{\rm res}$ should exhibit the same
scatter of about $0.4$ dex observed in $\Gamma_0$ (e.g., Beifiori et al.
2012; McConnell \& Ma 2013; Kormendy \& Ho 2013). This conclusion is quite
reasonable, since $\alpha_{\rm res}$ embodies a large number of physical
mechanisms and astrophysical settings, as we shall discuss in \S~5.2.
Actually this scatter is essential to reproduce the QSO luminosity functions
at very high redshift ($z\gtrsim 4$), that sample only $\gtrsim 3\,\sigma$
objects (Wyithe \& Loeb 2003; Mahmood et al. 2005; Lapi et al. 2006;
Fanidakis et al. 2012). All in all, the final mass of the BH itself can be
approximated as (see Fig.~3)
\begin{equation}
M_{\rm BH}\approx 5\times 10^{7}\, \alpha_{\rm res}\,
\left({M_{\rm H}\over 10^{12}\, M_{\odot}}\right)\, \left({1+z\over
3.5}\right)^{0.6}\, M_{\odot}~.
\end{equation}

In the original formulation by Granato et al. (2004) the star formation and
disc accretion are abruptly shut off when the mechanical energy delivered by
the active nucleus is large enough to unbind the gas from the galaxy. From
Eqs.~(2) and (3) we can infer that this occurs at a galactic age $t_{\rm
QSO}$ when the energy feedback from the QSO overwhelms that from SNae; such a
time is found to be very close to $\Delta t_{\rm burst}$ as defined below
Eq.~(4). Instead, in the present work we allow for a more educated decline,
that we will show to be essential in reproducing the observational data (see
\S~3.2). As a matter of fact, the feedback can require some time in order to
become fully effective. For instance, King (2010) has shown that the
expansion of a momentum-driven shell can reach $20$ kpc in a massive galaxy
on a timescale of about $6\times 10^7$ yr, similar to the typical $e$-folding
time $\tau_{\rm ef}\approx 4\times 10^7$ yr. Thus in the model we assume that
at the galactic time $t_{\rm decl}=t_{\rm QSO}+n\,\tau_{\rm ef}$ with
$n\approx 2$ both the ISM and the reservoir are affected by the QSO feedback.

For simplicity, we assume that the SFR fades on a timescale $\tau_{\rm SFR}$
and the accretion rate decreases on a timescale $\tau_{\rm AGN}$, free
parameters to be set by comparison with the data in \S~3. We shall express
them in terms of the timescale $\tau_{\rm ef}$ for BH growth. The
AGN declining phase lasts until the gas mass piled up in the reservoir is
exhausted. On the other hand, it does not occur if the gas in the reservoir
is exhausted in the time interval between $t_{\rm QSO}$ and $t_{\rm decl}$.
Presence or absence of the AGN declining phase depends on the parameter
$\alpha_{\rm res}$ following Eq.~(7), i.e., on the efficiency of the gas 
storing
within the reservoir; the transition between these two regimes occurs
around $\alpha_{\rm res}\approx 5$. Examples of the AGN light-curves are
plotted in Fig.~1.

\section{The relationship between star formation and BH growth: data and 
model
comparison}

The coevolution of AGNs and of their host galaxies at high redshift is mainly
traced by observations aimed at estimating the relationship between the star
formation in the host and the nuclear activity due to accretion onto the
central BH. There are two main ways to proceed. In the first, galaxies are
selected on the basis of their SFR and followed up with X-ray observations
looking for nuclear activity. In the second, the AGN is selected in X-rays
and then followed up with observations looking for star formation in the
host. The main outcomes of these observations are statistics on detections
and on properties such as stellar mass in the host and BH mass. These
statistics depend on the limiting flux/luminosity of the selection and of the
ensuing follow-up. Details on model estimates are given in the Appendix.

\subsection{The relationship between SFR and AGN luminosity in high redshift
starbursting galaxies}

We begin with discussing X-ray observations of high redshift galaxies mainly
selected on the basis of their SFR. The purpose is to illustrate how powerful
are accurate statistics derived from FIR and X-ray observations in
constraining the evolution of the star formation in the host and of the BH
growth, using the model lightcurves presented above as a guidance. The
available statistics already strongly support the conclusion that in an early
phase the AGN luminosity is increasing exponentially, while the SFR is almost
constant.

\subsubsection{FIR and (sub-)mm galaxies in X-rays: observations vs. model 
predictions}

In Fig.~5 we present the fraction of AGNs detected in X-rays ($2-10$ keV
band) as a function of the host FIR luminosity and of the X-ray detection
threshold. We warn the reader that for different data samples the minimum
X-ray luminosity for detection is somewhat uncertain (depending on flux
limit, absorption correction, redshift, and so on). The model agrees with the
data in predicting an increasing detection fraction with increasing SFR (or
$L_{\rm FIR}$) at given X-ray detection threshold $L_X$.

In the model the FIR luminosity is mainly a function of the halo mass (see
Eq.~4) and changes by a factor of $\lesssim 3$ for most $\gtrsim 80\%$ of
the star formation duration (see Fig.~1). On the other hand, the accretion
luminosity is increasing exponentially on a timescale $\tau_{\rm ef}$. As a
consequence, the time lapse during which the central AGN is brighter than the
X-ray detection threshold is only a fraction of the starburst duration. The
increasing fraction of detected AGNs with increasing SFR of the host galaxy
reflects the higher power that can be attained by more massive BHs in more
massive galaxy halos (see Eq.~9). Plainly, at given halo mass, the higher is
the X-ray detection threshold, the shorter is the fraction of time spent by
the AGN above that limit. The model also predicts a slight redshift
dependence, due to the combined effect of the increase in Eddington ratio
$\lambda$ (implying a decrease of $\tau_{\rm ef}$) and of the
decrease in the star formation duration with increasing redshift.

Conversely, a shorter or a longer timescale for the evolution of the
accretion luminosity would imply detection rates higher or lower than
observed, respectively. For instance, in the hypothesis that the luminous
accretion rate during the star formation epoch is proportional to the
large-scale SFR (e.g., Mullaney et al. 2012a; Silk 2013), $\dot{M}_{\rm
accr}\approx \Gamma_0\,\dot{M}_{\star}$ with $\Gamma_0\approx 3\times
10^{-3}$, we get $L_{\rm FIR}/L_{\rm AGN}\approx 1\, (\epsilon/0.15)^{-1}$.
Then the expected X-ray luminosity would exceed $L_X\approx 10^{42}\,
(k_X/25)^{-1}\, (\Gamma_0/3\times 10^{-3})\, (\dot
M_\star/M_{\odot}\,\mathrm{yr}^{-1})$ erg s$^{-1}\gtrsim 10^{43}$ erg
s$^{-1}$ whenever $\dot{M}_{\star}\gtrsim 10\, M_{\odot}$ yr$^{-1}$ and
$L_{\rm FIR}\gtrsim 3\times 10^{44}$ erg s$^{-1}$, adopting a standard
bolometric correction (e.g., Hopkins et al. 2007). Such an AGN X-ray
luminosity would imply that in galaxies with $L_{\rm FIR}\gtrsim 3\times
10^{44}$ erg s$^{-1}$ the nuclear activity is always detected, in contrast
with the observations presented in Fig.~5.

To escape the limits set by observations, one has to assume that all AGNs
hosted by FIR bright galaxies have extreme X-ray absorbing column densities
($N_{\rm H}\gg 10^{24}$ cm$^{-2}$). This would yield a selection bias which
allows us to detect only the objects with lower column densities, implying a
reduced detection fraction. Although quite large column densities ($N_{\rm
H}\gtrsim 3\times 10^{23}$ cm$^{-2}$) have been observed in several (sub-)mm
selected galaxies, a significant fraction of objects exhibits lower column
densities (see Alexander et al. 2005; Laird et al. 2010; Georgantopoulos et
al. 2011; Rafferty et al. 2011); in particular, no correlation between FIR
luminosity and column density has been observed.

The FIR/submm selection elicits galaxies with bolometric luminosity
associated to star formation larger than the AGN bolometric luminosity
($L_{\rm FIR}/L_{\rm AGN}\gtrsim 1$; see Alexander 2005; Georgantopoulos et
al. 2011; Laird et al. 2010; Rafferty et al. 2011; Johnson et al. 2013; Wang
et al. 2013). The few objects that exhibit a ratio $L_{\rm FIR}/L_{\rm
AGN}\sim 1$ have a large X-ray power $L_X\gtrsim 3 \times 10^{44}$ erg
s$^{-1}$ and large luminosity in the H${\alpha}$ line (Alexander et al.
2008). As for the model predictions, the AGN bolometric luminosity approaches
and possibly exceeds the galaxy FIR luminosity only for a couple of
$e$-folding times before its maximum (see Fig.~1). In addition, to get
$L_X\gtrsim 10^{44}$ erg s$^{-1}$, the minimu X-ray luminosity often assumed
to define X-ray QSOs, a value $\dot{M}_{\star}\gtrsim 100\, M_{\odot}$
yr$^{-1}$ or correspondingly $L_{\rm FIR}\gtrsim 3\times 10^{45}$ erg
s$^{-1}$ is required.

The claim of having detected stacked X-ray emission from Lyman Break Galaxies
(LBGs) at redshift $z\gtrsim 6$ by Treister et al. (2011) has been shown to
be uncorrect, being mainly due to problems in the background subtraction
(Willott 2011; Fiore et al. 2012b). Basu-Zych et al. (2013) pointed out that
the contribution from LBGs at $z\gtrsim 6$ to the X-ray background is only
minor. As can be seen from their UV restframe luminosity function (e.g.,
Bouwens et al. 2011; McLure et al. 2013; Oesch et al. 2013), most LBGs at
$z\gtrsim 6$ exhibit in fact $\dot{M}_{\star}\lesssim 2-5\, M_{\odot}$
yr$^{-1}$. In the model (see Mao et al. 2007) these SFRs correspond to halos
with $M_{\rm H}\lesssim 5\times 10^{10}\, M_\odot$, wherein the BH can only
acquire a relatively small mass $M_{\rm BH}\lesssim 10^5\, M_{\odot}$.

\subsubsection{$K$-band selected starforming galaxies}

The sample analyzed by Mullaney et al. (2012a), primarily selected in the
$K$-band (with $BzK$ color criteria chosen to ensure presence of star
formation), allows us to introduce the galaxy mass in the study of the
relationship between star formation and AGN activity. In the following we
focus on the sample at $z\approx 2$, relevant for our work. Because of the
spectral coverage from the UV to $24\, \mu$m and low detection limits, the
star formation can be probed down to small rates. The AGN activity is
inferred from X-ray observations with $4$ Ms exposure time, thus reaching
detection thresholds $L_X\approx 10^{42}$ erg s$^{-1}$ at $z\approx 2$. Such
a luminosity corresponds to central BHs of relatively low mass $M_{\rm
BH}\sim 10^5-10^6\, M_\odot$.

The detected AGN fraction as a function of FIR luminosity for the sample of
Mullaney et al. (2012a) is very similar to the results obtained for FIR or
(sub-)mm selected galaxies (see also Fig.~5), implying that the primary
$K$-band selection is not introducing a relevant bias with respect to a pure
FIR selection. Moreover, the average ratio of FIR to AGN bolometric
luminosity is large ($4\lesssim L_{\rm FIR}/L_{\rm AGN}\lesssim 9$), as found
for (sub-)mm selected galaxies.

The results of Mullaney et al. (2012a) show that the AGN detection fraction
is increasing with the stellar mass (see Fig.~6); the model reproduces this
behavior. However, these results are not independent of those presented in
Fig.~5. This is illustrated in Fig.~7, where data show an approximately
linear dependence of stellar mass on FIR luminosity or, equivalently, on SFR.
In the latter figure the solid black line shows the relation
$M_{\star}-L_{\rm FIR}$ between the average final stellar mass and the FIR
luminosity predicted by the model for a sample of star forming galaxies at
redshift $z\approx 2$, independently of the nuclear activity; this relation
can be understood on the basis of Eqs.~(4) and (5). We notice that there is
very weak or no dependence on the X-ray detection threshold, since the
relationship is dictated only by the star formation history. The change of
the $M_{\star}-L_{\rm FIR}$ relation with the redshift, apparent in Fig.~7
stems from the different dependence of mass and SFR on $z$. This also
explains why the fraction of detected AGNs at fixed stellar mass is almost
redshift independent (see Fig.~6) while the fraction at fixed FIR luminosity
decreases with decreasing $z$ (see Fig.~5).

Fig.~8 illustrates the predictions of the model for the ratio $\langle
L_X\rangle/\langle L_{\rm FIR}\rangle $ between the average X-ray and FIR
luminosities of FIR-selected galaxies as a function of the stellar mass. We
compare the model outcome with the data by Mullaney et al. (2012a). Though
the latter refer to both detected and undetected sources, the authors point
out that the X-ray counts for each bin in mass are dominated ($\gtrsim 80\%$)
by X-ray detected galaxies. For masses large enough to allow detection, the
ratio keeps almost constant.

The model reproduces such a behavior despite it features in individual
galaxies a SFR almost constant and a BH accretion rate exponentially
increasing. This comes from the primary selection in the $K$-band, which
biases toward late times of the evolution, where $L_{\rm AGN}$ is rapidly
approaching $L_{\rm FIR}$. Specifically, from the data it is apparent that
the AGN bolometric luminosity (with $k_X\approx 40$ for reference; see
Hopkins et al. 2007) is still a factor of several below $L_{\rm FIR}$, and it
will take another couple of $e$-folding time to attain it.

In Fig.~9 we show the ratio $L_{\rm FIR}/L_{\rm AGN}$, the specific SFR
(sSFR) $\dot M_\star/M_\star$ and the specific BH accretion rate
$\dot{M}_{\rm BH}/M_{\rm BH}$ as a function of time as predicted by the model
for a starforming galaxy at redshift $z\approx 2$; we focus on a halo mass
$M_{\rm H}\approx 2\times 10^{12}\, M_\odot$, corresponding to a stellar mass
of $M_\star\approx 6\times 10^{10}\, M_{\odot}$, typical of the sample
considered by Mullaney et al. (2012a). The range of the observed values of
the sSFR and of the luminosity ratio are marked on the respective model
predictions. It is apparent that the data on the luminosity ratio single out
a galaxy age $\approx 9-11\, \tau_{\rm ef}$, corresponding to $70-80\%$ of
the star formation duration. In terms of sSFR the time spanned by the data is
larger, possibly due to the larger uncertainties in the mass estimates.

Therefore, the selection of Mullaney et al. (2012a) picks out objects that
are on the average detected about $2-3$ $e$-folding times before the peak of
the AGN activity; as a consequence, they should exhibit a BH to stellar mass
ratio lower than the local one by a factor $5-10$. This is close to the
estimate obtained by Alexander et al. (2008) for a sample of (sub-)mm
selected galaxies exhibiting nuclear activity. Relatedly, in the model the
specific BH accretion rate stays constant at values around $10$ Gyr$^{-1}$,
mirroring the exponential time behavior of the BH accretion.

In conclusion, the data of Mullaney et al. (2012a) concur with the
FIR-selected samples in supporting the view that the BH and stellar mass are
assembling in parallel, but on different timescales. While the stellar mass
increases almost linearly with galactic age for a period $\Delta t_{\rm
burst}\approx 7 \times 10^8$ yr at $z\approx 2.5$, the BH mass is increasing
exponentially on a timescale $\sim \Delta t_{\rm burst}/15$. Such a timescale
indicates that during this epoch the BH accretion is in a self-regulated
regime. In \S~4.1 we shall see that this conclusion is also supported by the
estimates of the central BH mass in FIR and (sub-)mm selected galaxies. More
detailed statistics on the nuclear activity in FIR selected galaxies would
allow us to determine with high precision the ratio between the duration of
the star formation and the $e$-folding timescale of the BH accretion.

\subsection{Star formation in high redshift AGNs: the quenching
of the star formation and of the nuclear activity}

The data discussed above allowed us to reconstruct the growth phase of the BH
mass and of the nuclear activity, when star formation in the host galaxy is
still significant. We recall that such epoch is characterized by objects with
$L_{\rm FIR}/L_{\rm AGN}\gtrsim 1$.

Here we show that determinations of the SFR in high-$z$, X-ray or optically
selected AGNs, allow us to explore the phases when the AGN/QSO approaches its
maximum mass/luminosity and the quenching of star formation and of nuclear
activity sets in. Our analysis focuses on what can be learnt on these
quenching timescales from (i) FIR and (sub-)mm observations of X-ray selected
AGNs and QSOs; (ii) star formation in optically selected QSOs.

\subsubsection{Star formation in high redshift X-ray selected AGNs and QSOs}

In Fig.~10 we present the results relative to the statistics of the X-ray
selected AGNs whose host galaxy has been detected in the FIR/(sub-)mm.
Although some caution must be used in interpreting data obtained with
different luminosity threshold in the FIR, nevertheless it is apparent that
only a quite small fraction of powerful X-ray AGNs with $L_X\gtrsim 10^{44}$
erg s$^{-1}$ are detected in the FIR. Such a small fraction suggests that the
most luminous AGNs extend their X-ray emission beyond the epoch of constant
and large SFR into a phase when the star formation has been quenched more
rapidly than the nuclear activity.

In the model this later phase is present only if the reservoir is not
exhausted before the feedback quenches the star formation; in such cases, the
accretion can continue for some additional time. The mass in the reservoir is
determined by the efficiency of the mechanisms capable of transfering the
angular momentum of the cold gas involved in the star formation process; in
the model this aspect is encapsulated into the parameter $\alpha_{\rm res}$
in Eq.~(5). On the other hand, the distribution of $\alpha_{\rm res}$
reflects into the distribution of the BH to stellar mass ratio
$\Gamma_0=M_{\rm BH}/M_{\star}$; we recall that local samples suggest
$\Gamma_0\approx 3\times 10^{-3}$ with a small, but positive dependence on
the stellar mass and with $1\, \sigma$ spread $\approx 0.4$ dex.

Allowing for a Gaussian distribution of $\alpha_{\rm res}$ around the average
value $2$ with a dispersion of $0.4$ dex, we find that the bright tail
($L_X\gtrsim 10^{44}$ erg s$^{-1}$) of the luminosity function of X-ray
selected AGNs is dominated by sources with $\alpha_{\rm res}\gtrsim 5$ (see
also Fig.~4 in Lapi et al. 2006). Such value is used in the computation of
the lightcurves presented in Figs.~1 and 3. By further setting $\tau_{\rm
AGN}=2\times \tau_{\rm ef}$ and $\tau_{\rm SFR}$=$\tau_{\rm ef}/3$ we obtain
a good representation of the available statistics on AGNs detected in the FIR
(see Fig.~10). Note that the SFR quenching timescale $\tau_{\rm SFR}\lesssim
10^{7}$ yr is consistent with the estimate by Daddi et al. (2007) and
Maiolino et al. (2012).

The AGN light curves as a function of the galactic time are then exploited to
estimate the luminosity function at different redshift, properly taking into
account the halo formation rate (see Appendix for details). The results shown
in Fig.~11 indicate a very good agreement with the data at $z=2$ and $3$.

We stress that moving toward lower luminosities the typical values of
$\alpha_{\rm res}$ approach the mean value $\approx 2$, so that these objects
do not feature a prolonged declining phase. If they did, the low luminosity
tails of the luminosity functions would exceed observational determinations
(see also the discussion by Granato et al. 2006). Similar constraints also
come from statistical matching arguments between the AGN accreted mass
function and the local BH mass functions (see Yu \& Lu 2004).

In the top panels of Fig.~12 the effect of changing the timescale for the
decline of the SFR and of the AGN luminosity is illustrated. Keeping the same
timescale for the star formation decline but reducing that of the nuclear
activity would imply a much higher fraction of detected AGNs. The opposite is
true when prolonging the duration of the nuclear activity. By the same token,
if the star formation is switched off too early with respect to the AGN
accretion, then the detection fraction significantly decreases, while it
increases if significant star formation occurs during the last phases of AGN
activity. The middle left panel of Fig.~12 shows that the uncertainty in the
bolometric correction $k_X$ is a minor issue. However, it is instead relevant
for reproducing the X-ray luminosity function (cf. Fig.~11; see also Lapi et
al. 2006). The middle right panel elucidates the effect of changing the FIR
luminosity threshold. We stress that the thresholds span a factor of $10$, a
wide range even compared with the expected uncertainty associated to the
observational data. An uncertainty restricted to a factor of $2$ is
reasonably achievable with accurate analysis of the present data, and would
allow us to derive firm conclusions on the way the star formation and the
nuclear activity are switched off. Indeed, the figure highlights the
potential of accurate follow-up at FIR/(sub-)mm restframe wavelengths of
high-$z$, X-ray selected AGNs. The bottom left panel shows the effects of
varying the parameter $\alpha_{\rm res}$ and the final BH mass; increasing
$\alpha_{\rm res}$ causes more objects to feature the declining phase in the
nuclear activity, and implies a smaller fraction of sources detected in the
FIR-bright phase; the opposite holds when reducing $\alpha_{\rm res}$. The
bottom right panel illustrates the effects of varying the time when the
decline of the star formation and nuclear activity sets in. Increasing this
time makes the FIR-bright phase more prolonged, so raising the detection
fraction, and viceversa.

The observed average FIR luminosity of the AGN hosts as a function of their
X-ray luminosity is an additional test for the decline of the star formation
and of the nuclear activity (see Fig.~13). We stress that the difference in
redshift apparent in the model predictions is almost entirely due to
different thresholds at different $z$, that reflect the data sampling. The
statistics of these averages depend on the flux/luminosity distribution of
detected and undetected sources around the X-ray detection threshold. It is
clear that the X-ray primary selection picks out AGNs with $L_{\rm
FIR}/L_{\rm AGN}\gtrsim 1$ for $L_X\lesssim 10^{44}$ erg s$^{-1}$, while
ratios $L_{\rm FIR}/L_{\rm AGN}\lesssim 1$ are typical at larger X-ray
luminosity. Though caution is mandatory due to different detection limits,
this behavior is nevertheless consistent with the fact that the declining
phase of the X-ray luminosity is present on average in AGNs with $L_X\gtrsim
10^{44}$ erg s$^{-1}$, i.e., in massive galaxies endowed with substantial BH
masses. Meanwhile, in AGNs with $L_X\lesssim 10^{44}$ erg s$^{-1}$ the turn
off of the AGN activity occurs on a timescale of the same order or even
smaller than the turn off in the star formation. All in all, the data suggest
that the decline sets up at around $L_{\rm FIR}\sim L_{\rm AGN}$.

The faster decline of the star formation in luminous X-ray AGNs is also
supported by the results of Page et al. (2012), who claimed the detection of
a strong reduction in the SFR at increasing X-ray luminosity, by stacking AGN
positions with flux derived from \textsl{Herschel} Multi-tiered Extragalactic
Survey (HerMES) maps. This finding is not fully confirmed by Harrison et al.
(2012), who remarks that within the error bars it is possible that the SFR
keeps constant with increasing X-ray luminosity. However, their suggestion
depends on an upper limit to the SFR at $L_X\approx 3\times 10^{44}$ erg
s$^{-1}$, larger by a factor of $3$ with respect to that in Page et al.
(2012).

It is worth noticing that, as pointed out by several authors (see Harrison et
al. 2012; Mullaney et al. 2012b; Rosario et al. 2012), at high luminosities a
contribution from the AGN could add to the power from the star formation. The
amount depends on the wavelength of observation, since the AGN emission is
less important at (sub-)mm than at FIR restframe wavelengths (Richards et al.
2006; Polletta et al. 2007; Cai et al. 2013).

Exploiting optical and near-IR photometry, Mullaney et al. (2012b) estimated
the stellar mass in the host galaxy of AGNs with luminosity $10^{42} \lesssim
L_X\lesssim 10^{44}$ erg s$^{-1}$. They concluded that the FIR--detected host
galaxies exhibit average stellar masses $M_\star\approx 5 \times 10^{10}\,
M_{\odot}$, increasing by a factor of $\sim 2$ when the X-ray luminosity
changes by about two order of magnitudes, and possibly tracing a power law
$M_{\star}\propto L_X^{1/7}$ relationship. In the model the exponential
growth of the AGN luminosity, coupled with the almost constant SFR, implies
that the stellar mass depends only very weakly on $L_{\rm AGN}$ according to
\begin{equation}
M_{\star}(t) \propto \dot M_\star\times \ln L_{\rm AGN}(t)~.
\end{equation}
If mass estimates are available, the sSFR of the galaxies can be
investigated. In Fig.~14 it is apparent that the model predicts a decrease in
the specific star formation rate at large X-ray luminosities, $L_X\gtrsim
10^{44}$ erg s$^{-1}$. This is a straightforward consequence of the decline
in the SFR (see Fig.~13) during the final phase of the AGN and galaxy
coevolution, when the stellar mass is already piled up. The present data do
not show a clear trend. In particular we notice that the lack of data at high
X-ray luminosity, $L_X\gtrsim 5\times 10^{44}$ erg s$^{-1}$, depends on the
limited volume covered by current surveys (see Fig.~6 in Rovilos et al. 
2012).

All studies of X-ray selected AGNs agree on very weak or no correlation
between the X-ray absorbing column density and the FIR luminosity (e.g.,
Stevens et al. 2005; Shao et al. 2010; Lutz et al. 2010; Rosario et al. 2012;
Rovilos et al. 2012). This result suggests that most of the X-ray absorption
may originate not from the gas and dust directly involved in the star
formation process, but be more related to the very central regions (see the
review by Turner \& Miller 2009). The same conclusion has been indicated by
direct studies of $z\sim 2$ heavily absorbed QSOs (Page et al. 2011).

On the other hand, Brusa et al. (2010) show that the fraction of obscured
($N_{\rm H}\gtrsim 10^{22}$ cm$^{-2}$) AGNs strongly decreases with
increasing X-ray luminosity, and ranges from $\sim 80-90\%$ at $L_X\sim
10^{42-43}$ erg s$^{-1}$ to $\sim 10-20\%$ at $L_X\gtrsim 6 \times 10^{45}$
erg s$^{-1}$, with an abrupt transition around $L_X\sim 10^{44}$ erg
s$^{-1}$. For X-ray selected AGNs this luminosity corresponds, on average,
to the transition from bolometric luminosity dominated by star formation to
that dominated by nuclear activity (see Fig.~13).

These findings can be understood recalling that an unbiased X-ray selection
can pick up objects either before or after the SFR quenching. Specifically,
in the early epoch dominated by star formation (see Fig.~1) the gas is
abundant both in the reservoir and in the central regions of the galaxy, and
both settings can contribute to the obscuration. When the AGN reaches its
maximum power, the ISM is rapidly blown away by winds on a timescale
$\sim \tau_{\rm ef}/3$ and the possible obscuration due to the ISM 
drastically
decreases. On the other hand, the QSO winds may also affect the
reservoir/proto-torus and the innermost X-ray absorbing regions, for instance
peeling off the gas at higher latitude and reducing the covering factors.

Interestingly, the high fraction $\sim 40\%$ of FIR detections in a sample of
highly absorbed X-ray QSOs with $N_{\rm H}\gtrsim 10^{22}$ cm$^{-2}$ and
$L_X\gtrsim 10^{45}$ erg s$^{-1}$ found at $z\sim 1.5-2$ by Stevens et al.
(2005) can be understood on the same grounds. Since these authors select
\emph{highly obscured} QSOs, they on the average pick up objects that are
still retaining a large reservoir and a large amount of gas in their ISM, and
so are likely retaining a large SFR. Taking into account that the space
density of absorbed QSOs is $\sim 15\%$ of that of the unabsorbed ones at
given X-ray luminosity (see Page et al. 2011), the $\sim 40\%$ fraction of
FIR detections for highly absorbed QSOs is still consistent with the low
fraction $\lesssim 10\%$ for an unbiased selection at large X-ray luminosity
(Rosario et al. 2012; Page et al. 2012; Harrison et al. 2012).

The results of Stevens et al. (2005) are also supported by Mainieri et al.
(2011) and by Carrera et al. (2011, 2013), who have shown that a major
fraction of all obscured QSO hosts at $z\gtrsim 1$ are forming stars at
significant rates. More quantitatively, the absorbed X-ray selected QSOs
exhibit luminosity ratios in the range $0.3\lesssim L_{\rm FIR}/L_{\rm
AGN}\lesssim 5$ (Stevens et al 2005; Vignali et al. 2009; Gilli et al. 2011;
Feruglio et al. 2011), supporting the view that they typically reach their
large X-ray luminosity when star formation is still vigorous.

Inspecting Fig.~15, it is apparent that for obscured X-ray QSOs the observed
ratios $L_{\rm FIR}/L_{\rm AGN}$ correspond to an interval in time $\sim
3\,\tau_{\rm ef}$ before the AGN decline phase sets in (see cyan box). Since
during this epoch both the reservoir and the ISM are rich in gas and dust,
X-ray obscured QSOs are also obscured in the UV-optical bands and, as a
consequence, are on the average extremely `red' (e.g., Fiore et al. 2009).
After the onset of the QSO winds the SFR rapidly decreases and, as mentioned
above, the covering factor of the reservoir also decreases. In fact, the 
population
of X-ray unobscured QSOs exhibit $L_{\rm FIR}/L_{\rm AGN}\lesssim 0.3$ (e.g.,
Page et al. 2004) and they look similar to the pure optically selected QSOs
(see below).

In conclusion, the above considerations illustrate that the study of star
formation in well defined samples of X-ray selected AGNs can be extremely
useful in statistically defining the observational framework to which a
sensible theory of coevolution must conform. As a matter of fact, the X-ray
selected objects may lie not only on the rising branch of the X-ray
luminosity curve when the SFR is almost constant, but also in the declining
phase of the SFR and nuclear activity. The number of objects with $L_{\rm
FIR}/L_{\rm AGN}\gtrsim 1$ relative to that with $L_{\rm FIR}/L_{\rm
AGN}\lesssim 1$ is an important clue on the timescale of the SFR quenching by
the AGN feedback and on the timescale over which the nuclear activity itself
is switching off.

\subsubsection{Star formation in optically selected QSOs}

Star formation in some of the galaxies hosting an optically selected QSO has
been clearly detected in the (sub-)mm even at very high redshifts (see Omont
et al. 1996, 2001, 2003; Carilli et al. 2001; Priddey et al. 2003; Wang et
al. 2008; Serjeant et al. 2010; Bonfield et al. 2011; Mor et al. 2012), up to
$z\sim 7$ (Venemans et al. 2012). Most of these observations have targeted
QSOs endowed with very large bolometric luminosity $L_{\rm AGN}\approx
4\times 10^{47}$ erg s$^{-1}$.

Omont et al. (2001, 2003), Carilli et al. (2001), Priddey et al. (2003)
observed more than one hundred optically selected QSOs ($2\lesssim z\lesssim
4$) at mm wavelengths, with a detection rate $\approx 30\%$. At variance with
the X-ray absorbed QSOs, the optically selected ones exhibit $L_{\rm
FIR}/L_{\rm AGN}\lesssim 0.3$ with an average value $\sim 0.12$ for detected
sources. Millimeter flux stacking on undetected targets yields $\langle
L_{\rm FIR}\rangle/\langle L_{\rm AGN}\rangle \approx 1.5\times 10^{-2}$. The
typical average ratio for the detected and undetected QSOs is $\langle L_{\rm
FIR}\rangle/\langle L_{\rm AGN}\rangle \approx 0.06$. Mor et al. (2012)
detected in the sub-mm bands with \textsl{Herschel} five optically selected
QSOs at $z\sim 4.8$, deriving larger FIR to X-ray luminosity ratios $0.25
\lesssim L_{\rm FIR}/L_{\rm AGN}\lesssim 0.6$. However, as pointed out by
these authors, such values must be taken as upper limits, since the sub-mm
fluxes refer to regions within $10$ arcsec from the targeted QSOs and they
are close to or below the source confusion limit. Small ratios of FIR to
bolometric luminosity for high redshift QSOs are fully confirmed by the
results of the \textsl{Herschel}-ATLAS survey (see Serjeant et al. 2010;
Bonfield et al. 2011), which include also objects with lower bolometric
luminosity, $L_{\rm AGN}\sim 10^{46}$ erg s$^{-1}$.

The (sub-)mm searches have also been performed by targeting more than $60$
optically selected QSOs at $z\sim 6$, once more favoring quite powerful
sources, though there are already attempts to observe also QSOs fainter than
$L_{\rm AGN}\lesssim 10^{47}$ erg s$^{-1}$ (see Willott et al. 2007; Wang et
al. 2008, 2010, 2011; Omont et al. 2013; Leipski et al. 2013). The fraction
of detected \emph{luminous} QSOs with $L_{\rm AGN}\approx 3\times 10^{47}$
erg s$^{-1}$ turns out to be $\approx 30\%$, remarkably close to what is
found at lower redshifts. The average maximum ratio $\langle L_{\rm
FIR}\rangle/\langle L_{\rm AGN}\rangle \lesssim 0.3$, the minimum value
$\approx 0.02$ derived by stacking undetected sources, and the average value
$\approx 0.12$ for detected objects are quite close to those for lower
redshift samples.

Can our model cope with the results for bright QSOs? The (positive) answer is
in Fig.~16. We recall that in the model the bright QSOs are associated to
massive galaxy halos formed at high redshift, in agreement with the
clustering data (e.g., Hickox et al. 2011). If the optically bright QSOs on
average appear at a time $t_{\rm opt}\approx t_{\rm decl}$ when $L_{\rm
FIR}/L_{\rm AGN}\lesssim 0.3$ (only a handful of observed QSOs exceed this
limit), the subsequent evolution of the ratio predicted by the model is
\begin{equation}
{L_{\rm FIR}\over L_{\rm AGN}}(t\gtrsim t_{\rm opt})\approx {L_{\rm FIR}\over
L_{\rm AGN}}(t_{\rm opt})\, e^{-5/2\times (t-t_{\rm opt})/\tau_{\rm
ef}}\approx 0.3\,e^{-5/2\times (t-t_{\rm opt})/\tau_{\rm ef}}~.
\end{equation}
The time lapse between the larger ratio $L_{\rm FIR}/L_{\rm AGN}\approx 0.3$
and the ratio $L_{\rm FIR}/L_{\rm AGN}\approx 0.02$ obtained by stacking
non-detections amounts to $\approx 1.2\, \tau_{\rm ef}$. This is an estimate
of the time interval between the onset of the optical phase and the time when
bright QSOs are on the average no longer detected at mm wavelengths. Over
this time the optical luminosity decreases by a factor of $\approx 2$, while
the FIR luminosity decreases by a factor of about $30$. According to
Eq.~(11), the average time interval during which the bright QSOs are
detectable at mm wavelengths is given by $\sim 2/5\times \ln (0.3/0.12)\,
\tau_{\rm ef} \approx 0.36 \, \tau_{\rm ef}$, recalling that the
corresponding $L_{\rm FIR}/L_{\rm AGN}\approx 0.12$. As a result, the
expected detection fraction amounts to $0.36/1.2\approx 30\%$.

In this context, the similarity of the detected fraction ($\approx30\%$) of
bright QSOs at low and high redshift can be understood on the basis of the
above equation, i.e., ascribed to the similarity of the physical processes in
action at redshift $z\gtrsim 1.5$. Although the above estimates are somewhat
uncertain since the statistics depend on the FIR detection threshold,
nevertheless they highlight the potential of refined data on the ratio
$L_{\rm FIR}/L_{\rm AGN}$ to trace this phase of the bright QSO evolution.

These results refer to QSOs with bolometric luminosity $L_{\rm AGN}\gtrsim
6\times 10^{47}$ erg s$^{-1}$. However, the detection rate decreases to
$\lesssim 5-10\%$ when fainter QSOs with magnitude $M_{1450}\gtrsim -25$, or
$L_{\rm AGN}\lesssim 6\times 10^{46}$ erg s$^{-1}$ are considered, see Omont
et al. (2013). The same authors find $\langle L_{\rm FIR}\rangle/\langle
L_{\rm AGN}\rangle\approx 0.07$ for their observed $20$ QSOs, in agreement
with the findings for brighter ones. Willott et al. (2013) studied with
\textsl{ALMA} two \emph{faint} QSOs with $M_{1450}\gtrsim -25$; one of the
objects has been detected and features $L_{\rm FIR}/L_{\rm AGN}\approx 0.05$,
while the second one has not been detected, providing an upper limit $L_{\rm
FIR}/L_{\rm AGN}\lesssim 0.02$. Also Bonfield et al. (2011) find that the
fraction of detections decreases to $\approx 8\%$ when lower luminosity QSOs
($L_{\rm AGN}\sim 10^{45}$ erg s$^{-1}$) are included in the target sample.
In our framework two effects cooperate to get such results. First, at a given
final BH mass, the decrease of the nuclear luminosity corresponds to a faster
decrease of the SFR. Secondly, less luminous QSOs are on the average
associated with less massive BHs and less massive halos and, as a
consequence, with host galaxies exhibiting lower SFRs (see Fig.~16).

In Fig.~17 we show how the FIR detection fraction of optically selected QSOs
depends on the model parameters. In particular, we illustrate how it changes
when varying the timescales of the AGN (top left) and the SFR decline (top
right), the bolometric correction (bottom left), and the FIR detection
threshold (bottom right). The same comments to Fig.~12 hold, so we do not
repeat them here.

The observational data and the model suggest that the X-ray absorbed QSOs and
the optically selected QSOs are representative of two adjacent and subsequent
phases, encompassing the time when the AGN luminosity overcomes the FIR
luminosity of the host, while the QSO winds remove the ISM and reshape the
reservoir around the BH (see Fig.~15). The handful of optically selected QSOs
with $L_{\rm FIR}/L_{\rm AGN}\gtrsim 0.3$ can be understood as peculiar
objects, wherein the obscuration usually associated with large FIR to AGN
luminosity ratio does not preclude them to be selected in optical surveys.
For instance, the QSO J0129-0035 at $z=5.78$ is the faintest one found in the
SDSS and exhibits $L_{\rm FIR}/L_{\rm AGN}\approx 0.83$ (Wang et al. 2013).
It also features a quite weak Ly$\alpha$ line emission (Jiang et al. 2009),
as expected for a partially obscured object. On the other hand, we recall
that when $L_{\rm FIR}/L_{\rm AGN}\lesssim 0.3$ the AGN feedback is fully
removing the cold gas and stopping any flow toward the reservoir/torus, which
shrinks down in mass and size. Therefore, during this epoch the optical and
X-ray emissions are on the average less obscured (see \S~5). Thus the 
selected
objects appear as unobscured AGNs, exhibiting the well known correlation
between UV/Optical and X-ray emission (e.g., Lusso et al. 2010).

In our framework, luminous objects evolve along the following sequence: at
first they are galaxies violently forming stars and growing up a supermassive
BH in their centers. The BH manifests its mass accretion in hard X--rays
because the obscuration for these energetic photons is less dramatic. Then
the QSO winds clear up the ISM and an optical bright phase follows till the
exhaustion of the fuel in the reservoir (see Fig.~3). The sequence looks
similar to that suggested by Sanders et al. (1988), although in our model
major mergers are not the leading phenomenon. It is quite remarkable that
even at high redshift the stage of strong star formation in a QSO precursor
has been already detected. This is the case of HFSL3 at $z\approx 6.3$
(Riechers et al. 2013). Its gas mass $M_{\rm gas}\approx 10^{11}\,
M_{\odot}$, about $40\%$ of the dynamical mass, is distributed within a
radius $\lesssim 2$ kpc and its properties are consistent with our
scheme. Such system is predicted to evolve into a QSO like
SDSSJ114816.64+525150.3 at $z\approx 6.42$ within a few $e$-folding times. As
a matter of fact, in the latter object Walter et al. (2009) have detected a
huge star formation rate $\dot M_\star\approx 1700\,M_{\odot}$ yr$^{-1}$
within a radius $\lesssim 1$ kpc. We notice that SDSSJ114816.64+525150.3 has
a ratio $L_{\rm FIR}/L_{\rm AGN}\approx 0.2$, as expected in our framework
(see Fig.~15). An additional interesting aspect is that the FIR selected
galaxies exhibit larger abundance of low excitation gas with respect to the
QSOs (e.g., Riechers et al. 2011), again possibly marking a step in the
evolution.

From the above discussion we expect that the luminous optically selected QSOs
are detectable for $\approx 1-2$ $e$-folding times around the maximum of
their light curve. This visibility time is properly included in the
computations of the optical luminosity functions at various redshifts. The
results reported in Fig.~18 show a quite good agreement with the observations
at high redshifts.

An additional prediction of our overall picture is that the mass in
stars and in the BH are already settled to their final value in optically
selected QSOs, since both star formation and BH accretion are exponentially
declining during the optical bright phase. A discussion of this issue is
deferred to \S~5.1. In conclusion, though the AGN light curve around and
after the peak of the activity is modeled in order to fit the behavior of
X-ray selected AGNs, it nicely describes also the statistics of optically
selected QSOs.

\section{Additional observational constraints}

There are additional observations, concerning the evolution of the BH to
stellar mass ratio, the obscured to unobscured AGN fraction and the QSO
outflows that are less systematic because of their inherent difficulty, but
still provide useful constraints. In the following we discuss how the model
confronts with these observations.

\subsection{Stellar and the BH masses in QSOs}

As recalled in \S~1, the mass in the old stellar population correlates with
the mass of the central BH in local ETGs. On the other hand, it has been
claimed by many authors that the relationship changes with cosmic time. While
individual BH and stellar mass estimates for high redshift AGNs/QSOs must be
taken with caution, the data are nevertheless rather informative.

The first relevant piece of evidence is that the \emph{optically} selected
QSOs at $z\gtrsim 2$ exhibit on the average an $M_{\rm BH}/M_\star$ mass
ratio $3-20$ times higher than the local value, $\Gamma_0\approx 3\times
10^{-3}$, (e.g., McLure \& Dunlop 2004; Peng et al. 2006; Coppin et al. 2008;
Decarli et al. 2010; Merloni et al. 2010; Wang et al. 2010; Targett et al.
2012; Omont et al. 2013; Wang et al. 2013). The model envisages that such a
finding is due to a selection bias. In fact, the high luminosity tail
($L_{\rm AGN}\gtrsim 10^{47}$ erg s$^{-1}$) of the luminosity function of
optically selected AGNs/QSOs at high redshift $z\gtrsim 3$ comprises objects,
that are on the high side ($\gtrsim 3\,\sigma$) of the local $\Gamma$
distribution (e.g., Wyithe \& Loeb 2003; Mahmood et al. 2005; Lapi et al.
2006, see their Fig.~4; Lauer et al. 2007). Considering the scatter of $0.4$
dex of the local ratio, the values predicted by the model for optically
selected QSOs is $\Gamma\gtrsim 10^{-2}$, in good agreement with the
observational data.

Note that the model ascribes the large scatter in $\Gamma$ mainly to a
scatter in $\alpha_{\rm res}$, i.e., to variations in the amount of mass that
is first accumulated in the reservoir and then accreted onto the BH. In
addition, it is also possible that the average value of $\alpha_{\rm res}$
may be itself a function of halo mass and redshift (see Eq.~8). The model
predicts that optically selected QSOs have already assembled most of their
final stellar and BH masses. This does not exclude that some mass is
subsequently added, but the later evolution must have a minor impact on the
average.

The complementary view of the BH and stellar mass evolution is given by
(sub-)mm selected galaxies with detectable AGN activity. The results of Borys
et al. (2005), Alexander et al. (2005, 2008), Melbourne et al. (2011),
Carrera et al. (2011) strongly suggest that in these galaxies the ratio
$\Gamma\lesssim (0.1-0.3)\,\Gamma_0\approx 3-9\times 10^{-4}$ applies. This
is consistent with the view that in (sub-)mm selected galaxies the BH mass is
still piling up. The almost constant SFR coupled with the exponential
increase of the BH accretion rate makes this ratio increasing almost
exponentially before the declining phase. We stress that during this
self-regulated regime of the BH growth, the ratio $\Gamma$ does not depend on
the amount of matter stored in the reservoir, i.e., on $\alpha_{\rm res}$. In
relation to this point we recall that the stellar mass distribution estimated
by Mullaney et al. (2012b) in X-ray selected AGNs as a function of the X-ray
luminosity is well reproduced by the model.

\subsection{Fraction of obscured to unobscured AGNs/QSOs}

AGNs are mainly classified as obscured on the basis of emission line spectra,
X-ray absorption $N_{\rm H}\gtrsim 10^{23}$ cm$^{-2}$, and reddening
$E(B-V)$. Such a classification criteria have been recently applied by
several authors to objects selected in X--rays (see Bongiorno et al. 2012) or 
in the
near-IR (Glikman et al. 2007; Banerji et al. 2012, 2013), with the purpose of
understanding the fraction of obscured to unobscured AGNs/QSOs and of
searching for clues on the connection between the obscuration and the host
galaxy star formation activity. As usual, we are interested in high redshift
sources, $z\gtrsim 1-1.5$. The fraction of red QSOs has been found to be
significant. For instance Glikman et al. (2007) found that the fraction of
reddened QSOs with $K\lesssim 14$ can range from $25\%$ to $60\%$ of the
total underlying population. With different selection criteria on color and
limiting magnitude Banerji et al. (2012) found that at very bright
luminosities $M_i \lesssim -30$, the red QSOs are $\sim 5$ times more
numerous than the UV bright ones. However, just below that limit the fraction
decreases significantly to values close to those found by Glikman et al.
(2007).

On the basis of the model we expect that an obscured AGN phenomenology occurs
before the action of the QSO wind when $L_{\rm FIR}/L_{\rm AGN}\gtrsim 0.3$
and the UV-optical lines and continuum emission of the AGN are heavily
absorbed by the reservoir/torus and by the dust-rich ISM. Contrarily,
unobscured sources are more likely selected after the onset of the QSO wind,
when the ISM and the reservoir have been significantly impoverished. If we
assume that the phase singled out by the X-ray absorbed QSOs coincides with
that of the red QSOs, then on the basis of Fig.~1 and the discussion in
\S~3.2.2, a rough estimation of a $2.5:1$ proportion between red and
UV-selected QSOs can be derived.

Before the setting of QSO winds, during the obscured phase, the size of the
highly ionized region of the ISM surrounding the QSOs is limited, since most
of the UV and ionizing photons are heavily absorbed. The small size of this
region around ULASJ112001.48+064124.3 at $z\approx 7.1$ (the highest redshift
QSO detected so far) claimed by Mortlock et al. (2011) can be ascribed to
such an effect, as also supported by the radiative transfer simulations by
Bolton et al. (2011). These authors found that the UV bright phase should
have lasted for about $10^7$ yr. Since ULASJ112001.48+064124.3 exhibits
$L_{\rm FIR}/L_{\rm AGN}\approx 3\times 10^{-2}$, our model would predict a
UV bright phase duration of about one $e$-folding time, which in the model at
that redshift amounts to $\approx 2\times 10^7$ yr. The agreement is
reasonable, taking into account the uncertainties associated to the data
and modelization. Interestingly, this example suggests that the early, dusty
phase of QSOs at very high redshift can be easily probed by the \textsl{James
Webb Space Telescope} with the NIRCam and MIRI instruments (Dickinson 2013),
as it can be explored at lower redshift with the WISE All Sky Survey (Wright
et al. 2010; Banerji et al. 2013).

\subsection{Large scale QSO outflows}

The AGN feedback on the ISM should manifest itself in terms of large outflows 
(see
Fabian 2012 for a review), signalling the rapid expulsion of gas and the
dramatic shutoff of the star formation. Large outflows have been observed in
relatively nearby galaxies hosting an active nucleus (Feruglio et al. 2010;
Rupke \& Veileux 2011; Sturm et al. 2011; Greene et al. 2012). At larger
redshift $z\gtrsim 1$ QSOs winds associated with large mass outflows have
been detected by several studies (e.g., Maiolino et al. 2001, 2004; D'Odorico
et al. 2004). Optically thick gas moving with large velocity around QSOs up
to a distance of $100$ kpc has been detected by Prochaska \& Hennawi (2009),
possibly associated with quite massive outflows.

Broad-absorption line QSOs constitute the most basic and longstanding aspects
of outflow phenomenology; line-driven outflows are expected in the form of
winds that can form just above the accretion disk by a combination of
radiation and gas/magnetic pressure (see Zubovas \& King 2013). In fact,
massive outflows have been confirmed in such objects by many X-ray
observations (see Brandt \& Gallagher 2000; Chartas et al. 2003, 2009).
Recently, Borguet et al. (2013) detected the most energetic QSO outflow to
date with a kinetic power of $\sim 10^{46}$ erg s$^{-1}$ and an associated
mass flow rate of $\sim 400\, M_\odot$ yr$^{-1}$.

The most evident massive QSO outflow has been detected by Maiolino et al.
(2012) with extremely accurate and detailed observations of the C[II] line in
SDSSJ114816.64+525150.3 at $z=6.42$. These authors estimate an outflow rate
$\dot{M}_{\rm out}\approx 3000\, M_{\odot}$ yr$^{-1}$ capable of removing the
gas in the host galaxy within $6\times 10^6$ yr. Outflow rates predicted by
our model for massive BHs (see Fig.~19) range from several hundreds to
several thousands $M_{\odot}$ yr$^{-1}$; at given mass the rates depend also
on redshift, chiefly through the specific binding energy entering Eqs.~(2)
and (3).

The model also requires that before the peak of the nuclear activity there
are weaker galactic winds mainly driven by stellar feedback, as observed in
some starforming galaxies at substantial redshift (Alexander et al. 2010;
Genzel et al. 2012). Both AGN and stellar feedbacks can remove a lot of cold
gas from the central regions of galaxies, though on different timescales. For
instance in SDSSJ114816.64+525150.3 the dynamical mass within a $\sim 2.5$
kpc radius is about twice the gas mass (Wang et al. 2010). The results of
Maiolino et al. (2012) suggest that the central regions of the host are
deprived by about half of their mass in $\lesssim 10^7$ yr. So these outflows
must affect the inner structure of the host galaxies, as pointed out by
several authors (see Fan et al. 2008, 2010, 2013; Damjanov et al. 2009;
Zubovas et al. 2013).

It is worth mentioning that, while usually the AGN feedback is invoked to
turn off the star formation, the latter may occur in the outflowing shell
(see Ishibashi \& Fabian 2012; Silk 2013; Zubovas et al. 2013). Since FIR
selected samples show that the SFR is high even when the AGN luminosity (and
hence the feedback) is low, this mechanism can originate only a fraction of
the total stellar mass, while still relevant with respect to the size of the
host galaxy (Ishibashi \& Fabian 2012). In particular, in the early phases of
the evolution, the SN feedback is energetically larger than the AGN one, and
the observations reveal large stellar mass in the presence of large gas 
abundance
and weak AGN activity (i.e., low mass BH).

In summary, large outflows able to remove most of the ISM from AGN hosts are
detected at least in one case and possibly in several others. We recall that
even the short duration of the phase can significantly depress the statistics
of outflow detection. For instance, in the case of SDSSJ114816.64+525150.3
the gas within a scale of $5-10$ kpc is going to be removed in $\approx
6\times 10^6$ yr, which is $20\%$ of the duration of the QSO optical phase.

\section{Discussion}

Here we first discuss the observational evidence on, and properties of the
circumnuclear structures in local AGNs, related to the presence of a massive
gas reservoir as predicted by the model (see Fig.~3). Then by extrapolating
the results of local observations to higher redshifts, we show that
\textsl{ALMA} high resolution imaging and coordinated X-ray high resolution
observations of strongly lensed (sub-)mm selected galaxies can cast light on
the epoch of stellar and BH mass growth. In the light of the above scenario,
we point out several physical mechanisms operating on different scales that
can lead to the accretion of a fraction of the ISM.

\subsection{The reservoir}

As shown in Fig.~3 the model predicts that around the central BH a gas
reservoir soon sets up. Its formation stems from the requirement of
funnelling gas of low angular momentum at a rate proportional to the SFR,
while the BH accretes in a self-regulated regime. Such a structure could
constitute a (proto-)torus around the BH of the same kind as that observed at
low redshift and often called for in order to explain the diversity of
unobscured and obscured AGNs (see \S~4.2). This torus/reservoir, with size
ranging from a few to several tens of parsecs, has been studied in detail
only in nearby AGNs. High spatial resolution observations reveal that
molecular gas and dust is largely present (e.g., Muller-Sanchez et al. 2009;
Krips et al. 2011; Diamond-Stanic \& Rieke 2012; Sani et al. 2012; H\"{o}nig
et al. 2013), often accompanied by star formation (e.g., Davies et al. 2007;
Storchi-Bergmann et al. 2012). The dust distribution looks quite complex,
with the hot dust located in compact structures, while the warm one is more
diffuse (see H\"{o}nig et al. 2013).

The studies of Kawakatu et al. (2007), Spaans \& Meijerink (2008) and
Maiolino (2008) concluded that a reservoir/torus with mass $M_{\rm
res}\approx 10^8-10^9\, M_{\odot}$ extending to $\sim 100$ pc can be revealed
up to redshift $z\lesssim 2$ by \textsl{ALMA}, thanks to its extraordinary
sensitivity and exquisite spatial resolution. In fact, recently
\textsl{ALMA}, though in a still incomplete technical configuration, has
revealed CO, HCN, HCO$^+$ emission lines in the very central regions of a few
nearby objects (see Combes et al. 2013; Fathi et al. 2013; Izumi et al.
2013).

In high redshift galaxies the mass in cold gas/dust relative to that in stars
is expected to be much larger than in local AGNs. In fact, CO emission lines
have been detected by \textsl{ALMA} in a number of high redshift $z\sim 2-3$
objects (see Weiss et al. 2013). In addition, the potential efficiency of
\textsl{ALMA} for these studies is boosted up in the case of strongly lensed
(sub-)mm galaxies. As predicted by Negrello et al. (2007, 2010; see also
Blain et al. 2004) on the basis of the model, strongly lensed starforming
galaxies can be efficiently selected with large area, relatively shallow
(sub-)mm surveys.

In this context, Gonzalez Nuevo et al. (2012) have shown that about $10^3$
strongly lensed starforming galaxies at redshift $z\gtrsim 2$ can be easily
extracted from the \textsl{Herschel}-ATLAS survey that covers $\sim 550$
deg$^2$. About $5\%$ of them are magnified by a factor $\gtrsim 10$ (see Lapi
et al. 2012; Bussmann et al. 2013). Additional starforming, strongly lensed
galaxies have been detected by the \textsl{Herschel}/HerMES (see Wardlow et
al. 2013) and by the \textsl{SPT} survey of about $1200$ deg$^{-2}$ at mm
wavelengths (Vieira et al. 2013). Taking the Kawakatu et al. (2007, see their
Fig.~2) model as a reference, one finds that lensing by factors $\gtrsim 10$
brings the apparent size of a reservoir with physical size $R_{\rm
res}\approx 100$ pc to $\approx 0.07$ arcsecs (or $0.015$ arcsecs for a
$R_{\rm res}\approx 20$ pc physical size), well above the resolution
achievable with \textsl{ALMA} for almost all CO transition lines.
Specifically, the CO emission lines with $J\geq 2$ are detectable in $12$ hr
at $5\,\sigma$ for redshift $z\gtrsim 2$. CO(6-5) and CO(5-4) are detectable
in about $4$ hr (Kawakatu et al. 2007, see their Fig.~3). Even the HCN lines
can be observed at least for higher $J$ (Kawakatu et al. 2007, see their
Fig.~4). The C[II] $158\,\mu$m line can also be excited in the molecular
torus by a combination of the AGN emission and of the possible star formation
(P\'{e}rez-Beaupuits et al. 2011). If so, \textsl{ALMA} in its final
configuration will be able to produce detailed mapping of the C[II] emission
in the torus up to large redshifts.

We caution that the Kawakatu et al. (2007) model makes the classic assumption
of a smooth, homogeneous dust distribution (e.g., Pier \& Krolik 1992;
Granato \& Danese 1994; Efstathiou \& Rowan-Robinson 1995). Clumpy torus
models may be more realistic, as demonstrated by mid-IR observations of
silicate emission features in local AGNs (see Nenkova 2008a,b; H\"{o}nig et
al. 2010; Kawaguchi \& Mori 2011; also H\"{o}nig 2013).

Efstathiou et al. (2013) presented a detailed analysis on the spectrum of
IRAS 08572+3915, a nearby starburst galaxy endowed with an AGN. Combining new
\textsl{Herschel} observations with previous near- and far-IR datasets they
showed that the dusty torus illuminated by the AGN contributes about $90\%$
of the total IR luminosity. They also demonstrated that the far-IR luminosity
in the wavelength range $40\la \lambda \la 800\, \mu$m of a smooth torus is a
factor $\la 2.5$ lower than that of a clumpy torus. The measured total SED
falls in between the smooth and the clumpy torus SEDs (cf. their Fig.~3). All
in all, this result suggests that the the far-IR continuum emission of a
smooth torus can be taken as a lower limit to the case of a clumpy torus.

As for the high-J CO emission lines, Kawakatu et al. (2007) find their
results to agree with the outcomes from numerical simulations of 3-D,
non-local thermal equilibrium radiative transfer in inhomogeneous dusty tori
(see Wada \& Tomisaka 2005). In fact, recent observations of nearby Seyfert
galaxies with \textsl{Herschel} (see Hailey-Dunsheath et al. 2012; Spinoglio
et al. 2012; Pereira-Santaella et al. 2013) have revealed high-J CO emission
lines roughly consistent with the expectations from homogeneous torus models.
We also remark that mid-IR emission from clumpy tori, as modeled by H\"{o}nig
\& Kishimoto (2010) and investigated by means of high-spatial resolution
mid-IR spectro-photometry of Seyfert galaxies by H{\"o}nig et al. (2010), may
be detectable with the \textsl{James Webb Space Telescope} even in
high-redshift galaxies provided that they are gravitationally lensed at
magnifications $\gtrsim 10$ (see Kawakatu \& Ohsuga 2011).

Maps and velocity profiles of the molecular lines in strongly lensed
starbursting galaxies will produce invaluable information on chemical
composition, kinematic and mass of the reservoir during a crucial phase of
galaxy evolution and BH accretion. In particular instances of quite large
gravitational amplification we can even get estimates of the BH mass. These
observations should be complemented by X-ray observations, which for
magnification $\gtrsim 10$ will allow us to get a detailed view of the 
nuclear
emission even for highly obscured and quite low luminosity objects (Matt et
al. 2004; Georgantopoulos et al. 2011). The predicted intrinsic X-ray
luminosity for typical lensed galaxies exceeds $L_X\gtrsim 10^{42}$ erg
s$^{-1}$; the magnification by factors $\gtrsim 10$ makes them easily
detectable in pointed observations up to $z\gtrsim 3$ with \textsl{Chandra}
and the next generation of X-ray instruments.

In conclusion, coordinated, high-resolution \textsl{ALMA} and X-ray
observations offer the possibility to probe the reservoir/torus and the AGN
activity in many tens of lensed, FIR-selected galaxies. The outcome will
provide an enormous progress in understanding the formation and coevolution
of stars and BH in primeval galaxies.

\subsection{The gas path from the ISM to the accretion disc}

\subsubsection{From the ISM to the reservoir}

The first step for the ISM gas in its path toward the central BH is the
reservoir. The model assumes that, when the star formation is active in the
host galaxy, some mechanism is able to drive a fraction of the ISM gas into
the \emph{reservoir} at a rate proportional to the SFR, according to Eq.~(7).
Several physical mechanisms can cause a fraction of the gas in galaxies to
lose angular momentum and to pile up in the very central regions. A non
exhaustive list includes gas drag, dynamical friction of gas plus star
clumps, tidal fields, spiral waves, winds and bars, radiation drag (e.g.,
Norman \& Scoville 1988; Shlosman et al. 1989, 1990; Shlosman \& Noguchi
1993; Hernquist \& Mihos 1995; Noguchi 1999; Umemura 2001; Kawakatu \&
Umemura 2002; Kawakatu et al. 2003; Thompson et al. 2005; Bournaud et al.
2007, 2011; Hopkins \& Quataert 2010, 2011). In general the presence of
clumps, which may be generated by fragmentation of gas already organized in
an unstable disc or by inflow of gas and star subclumps, tends to increase
the efficiency of such mechanisms.

Since the loss of angular momentum is a requirement also for the cold gas to
collapse into clouds and to fragment into stars, the starburst activity and
the AGN fueling have been often associated. Gravitational torques acting on
gaseous and stellar disks can induce radial flows of the ISM (e.g., Shlosman
et al. 1989, 1990), thus increasing the central surface density, triggering
star formation in the inner kpc scale and moving gas toward the $100$ pc
region. Such gravitational torques can be induced by external events, such as
tidal encounters, orbital torques by satellites and minor/major mergers, or
by internal instabilities related to infall, bars, asymmetries generated
during the evolution of the gas plus star structure. Disc instabilities also
grow gaseous clumps that migrate toward the central regions by dynamical
friction (Shlosman \& Noguchi 1993). The clumps can then contribute to the
formation of the bulge (Noguchi 1999; Genzel et al. 2010, 2011; Immeli et al.
2004; Ceverino et al. 2010; Bournaud et al. 2007, 2014). All these processes
are expected to establish a kind of relationship between the rate of star
formation, at least in the inner kpc scale, and the rate at which the gas is
delivered toward the very central regions of the galaxy.

Hopkins \& Quataert (2010, 2011) have explored in detail with numerical
simulations the effects of gravitational torques (externally or internally
induced). They show that part of the ISM gas is shocked and dissipates energy
and angular momentum, flowing toward the central regions. On scales of $1-30$
pc such a gas can reach large surface densities, $\sim 10^{11}\, M_{\odot}$
kpc$^{-2}$. This gas can be identified with the reservoir/torus introduced in
our model.

In summary, there are several physical mechanisms that can plausibly reduce
the angular momentum at least for a fraction of the diffuse ISM gas and drive
it to migrate from kpc to $\lesssim 100$ pc scale at a rate proportional to
the SFR of the host galaxy. This possibility is encased into Eqs.~(7) and (8)
of the model. We stress that data support a large variance in the coefficient
$\alpha_{\rm res}(M_{\rm H},z)$ appearing there, which looks quite plausible
given the suggested mechanisms and the results of numerical simulations.

\subsubsection{From the reservoir to the accretion disc}

Once low angular momentum gas has accumulated in the reservoir, additional
loss/tranfer of angular momentum is required in order to bring the gas toward
the accretion disc at sub-parsec scales. Some of the mechanisms at work on
larger scales have been proposed to operate also on the smaller ones; for
instance, bars-in-bars instabilities (Shlosman et al. 1989, 1990),
gravitational interactions and dynamical friction in clumped disc (Kumar
1999) have been proposed.

Several authors have pointed out that in the circumnuclear regions on parsec
scales the gas is very likely rich in metals and dust, prone to fragment in
clumps and to form stars. On the one hand, the stellar feedback can remove
part of the gas from the reservoir; on the other hand, it can favor the gas
to flow toward the accretion disc (e.g., Wada \& Norman 2001, 2002; Thompson
et al. 2005; Murray et al. 2005; Kawakatu \& Wada 2008). For instance, in the
model proposed by Kawakatu \& Wada (2008) the gas turbulence supported by SN
explosions transports angular momentum. As a result, a fraction of $\sim
10-30\%$ of the gas flowing from larger scales can migrate toward the
accretion disc. Also in the numerical simulations by Hopkins \& Quataert
(2010) there is a correlation between SFR in the circumnuclear disc and the
rate of accretion onto the BH. Thompson et al. (2005) find two possible
classes of circumnuclear discs depending on the rate at which the gas is
supplied: discs with star formation large enough to consume most of the
available gas and with practically no accretion onto the central BH;
alternatively, discs with large star formation only at their periphery, still
leaving enough gas to ensure significant accretion.

Storchi-Bregmann et al. (2012) have detected star formation in NCG1068 on a
scale of $\sim 100$ pc. Sani et al. (2012) confirmed this result for $4$
additional nearby Seyfert galaxies, but concluded that in inner regions the
physical conditions of the gas are not favorable to star formation. On the
theoretical side, Begelman \& Shlosman (2009) pointed out that in the
circumnuclear discs highly turbulent continuous flows are relatively stable
against fragmentation.

A relevant point is that the star formation and its associated feedback in
the reservoir should \emph{not} be too much efficient; otherwise, the amount
of gas available to flow toward the BH will be substantially reduced and
eventually the relationship between stellar and BH mass will be erased.
Likely, in the balance the overall mass of long living stars formed in the
reservoir should be much less than the mass accreted onto the BH; this does
not imply that the star formation is not important in helping the BH feeding,
as mentioned above.

\subsubsection{The accretion rate and the effect of feedback}

In order to fit the observed statistics of both (sub-)mm/FIR and X-ray
selected objects at $z\gtrsim 2$, the  model assumes that the accretion is
self-regulated with a time-averaged Eddington ratio
$1\lesssim\lambda\lesssim 4$, depending on the redshift, and constant mass to
energy conversion efficiency $\epsilon\approx 0.15$. The model also envisages
that, when the AGN feedback becomes dominant and the reservoir is no more
fed, the accretion becomes supply-limited and sub-Eddington. These outcomes
can be compared to the predictions from theoretical treatments and numerical
simulations concerning the accretion onto the central BH.

Analytical treatments and hydrodynamical simulations have been used so far to
explore possible effects of the AGN radiative power on the accretion rate
(e.g., Begelman 1979; Abramowicz \& Lasota 1980; Abramowicz et al. 1988;
Watarai et al. 2000; Watarai 2006; Ohsuga 2007; Ohsuga \& Mineshige 2011;
Fabian 2012). In general, the rate $\dot{M}_{\rm accr}$ of gas accretion onto
the BH is possibly only a fraction of the rate $\dot{M}_{\rm inj}$ at which
the gas is supplied from the reservoir toward the BH, i.e., $\dot{M}_{\rm
accr}=f_{\rm accr}\dot{M}_{\rm inj}$ due to possible outflows (Begelman 2012;
Watarai 2006; Ohsuga 2007; Li 2012); the net effect is a kind of
self-regulation (see Debuhr et al. 2010, 2011, 2012). Depending on the local
physical conditions, accretion and outflow rates, mass-to-radiation
efficiencies and Eddington ratios are expected to fluctuate on very short
time scales (e.g., Ohsuga 2007; Li 2012); we stress that in the model we
refer to quantities averaged over the timescale needed by the BH to acquire
its final mass.

In the model it has been assumed that most of the matter delivered from the
reservoir toward the BH does accrete and only a negligible fraction goes into
winds, i.e. $f_{\rm accr}\approx 1$. Once more this is required in order not
to lose the correlation between the BH and stellar mass. However, a
significant variance in $f_{\rm accr}$ is expected and should produce a
variance in the final BH masses.

Accretion rate and radiation efficiency are coupled in accretion discs and
depend on the BH specific angular momentum (Abramowicz et al. 1988; Watarai
2006; Li 2012; Narayan et al. 2012; Roth et al. 2012; Li et al. 2013; Liu et
al. 2013). The possibility that a regime of photon trapping can infringe the
classical Eddington limit has been pointed out, together with the general
conditions for its onset and decay (e.g., Blandford \& Begelman 2004;
Begelman et al. 2006; Begelman \& Shoslman 2009; Begelman 2012).

Watarai (2006) presented useful approximations of numerical and analytical
results for thick/thin disks in the case of super/sub-Eddington accretion. If
the accretion rate is parameterized as $\xi=\dot{M}_{\rm accr}\, c^2/L_{\rm
Edd}$, then the Eddington ratio can be approximated by
\begin{equation}
\lambda={L\over L_{\rm Edd}}\simeq 2\, \ln\left[1+{\epsilon_0\over 2}\,
\xi\right]~,
\end{equation}
where $\epsilon_0\approx 0.3$ is the efficiency for a maximally rotating BH;
we recall that the spin of the BH rapidly increases to this maximal
value during super-Eddington accretion episodes. The radiative efficiency
during accretion is bound to be $\epsilon=\lambda/\xi$. As a consequence, the
accretion rate may exceed the Eddington one ($\xi> 1$), nevertheless keeping
the ratio $\lambda\lesssim 4$ and the radiative efficiency $\epsilon\lesssim
0.15$.

How does our basic model compare with this theoretical framework? In Fig.~20
the value of accretion rate $\xi$, Eddington ratio $\lambda$ and radiative
efficiency $\epsilon$ are presented as a function of galactic age for a QSO
shining at $z=6$ with a bolometric luminosity $L_{\rm AGN}\approx 3\times
10^{47}$ erg s$^{-1}$ and associated to a halo mass $M_{\rm H}\approx 2\times
10^{12}\, M_{\odot}$. The accretion rate implied by the model and the data is
almost constant with $\xi\approx 25$ for $\approx 15$ $e$-folding times and
then it decreases exponentially till the reservoir is exhausted. This opens
the theoretical issue of finding mechanisms able to provide accretion rates
exceeding the Eddington one, but up to such a value $\xi\approx 25$. Notice
that in this case $\tau_{\rm ef}\approx 1.8\times 10^7$ yr and the growth
phase lasts $3\times 10^8$ yr, less than the age of the Universe (amounting
to $\sim 1$ Gyr at $z\sim 6$). Watarai's approximation yields $\lambda\approx
3$ and $\epsilon\approx 0.13$, very similar to the values adopted in the
model (see Fig.~20). Similar results have been found by Li (2012), who has
also explored the possibility of having a largely super-Eddington regime
followed by a sub-Eddington one in order to explain the presence of the most
massive BH, $M_{\rm BH}\approx 10^9\, M_{\odot}$ at $z\approx 6$.

Figure~20 shows that the rapid decrease of the luminosity after the quasar 
wind
and throughout the quasar optical phase, entails a rapid decrease of
$\lambda$. This behavior can be compared with the Eddington ratios inferred
for optically selected QSOs. The observed distribution of Eddington ratios
$\lambda\equiv L_{\rm AGN}/L_{\rm Edd}$ in $z\approx 6$ QSOs estimated by
Willott et al. (2010) exhibits a lognormal distribution peaked at
$\lambda\approx 1$, with a dispersion $\approx 0.3$ dex; for comparison, the
distribution of $z\approx 2$ QSOs peaks at $\lambda\approx 0.37$. The trend
of decreasing Eddington ratio with redshift has been confirmed exploiting
quite large samples by Kelly \& Shen (2013), who also found that for optical
QSOs at $z\lesssim 5$ the Eddington ratio takes on a maximum value
$\lambda\approx 3$. In the model during the exponential growth of the BH the
maximum Eddington ratio is around $\lambda\approx 4$ at $z\approx 6$ and
$\lambda\approx 1$ at $z\approx 2$, and decreases exponentially from those
values during the optical bright phase (cf. Fig.~20), reproducing the
observed behavior.

We notice that with $\xi\lesssim 25$, the radiative efficiency
$\epsilon\approx 0.13$ is still large enough to have significant luminosity
even during the super-Eddington phase. On the contrary, a too small
efficiency would affect the fraction of nuclear activity detections in FIR
selected objects (see Fig.~10). During the optical bright phase the
efficiency $\epsilon=\lambda/\xi$ increases approaching the limit of $0.3$.
Unfortunately, observational determinations of the efficiency for individual
AGNs are still problematic (Raimundo et al. 2012). We recall that $\lambda$
and $\epsilon$ are key parameters in the estimate of the mass accreted onto
the supermassive BH during the bright AGN phase (e.g., Salucci et al. 1999;
Marconi et al. 2004; Shankar et al. 2004, 2013: Kelly \& Merloni 2012), and
average values $\epsilon\gtrsim 0.1$ are required by abundance matching
arguments.

After the ejection of the cold gas by the QSO feedback, the mass in the
reservoir, if not completely exhausted, is no longer sufficient to sustain a
super-Eddington accretion (the BH mass is large); thus a supply-limited,
sub-Eddington accretion regime sets in during the declining phase of the AGN
luminosity. We recall that the BH influence radius $\approx G\, M_{\rm
BH}/\sigma_\star^2$ itself increases exponentially before the peak of the AGN
activity, to attain values $\approx 70\, (M_{\rm BH}/10^9\, M_\odot)\,
(\sigma_\star/250\,\mathrm{km~s}^{-1})^{-2}$ pc, close to the possible
reservoir size. In these conditions, the standard theory of thin accretion
discs should apply. A naive estimate of the accretion rate reads
\begin{eqnarray}
\dot{M}_{\rm accr}={M_{\rm res}\over \tau_{\rm visc}}={\sigma_\star^3\over
G\, \mathcal{R}_{\rm crit}}\, {M_{\rm res}\over M_{\rm BH}}~;
\end{eqnarray}
here, according to standard prescriptions (e.g., Burkert \& Silk 2001;
Begelman 2012; King 2012), the viscous timescale $\tau_{\rm visc}\sim
\mathcal{R}_{\rm crit}\times \tau_{\rm dyn}$ is taken as the dynamical time
$\tau_{\rm dyn}\approx G M_{\rm BH}/\sigma_\star^2 \times 1/\sigma_\star$ at
the BH influence radius times the critical Reynolds number, $\mathcal{R}_{\rm
crit}\sim 10^{2-3}$, for the onset of turbulence.

However, the typical accretion rates derived from Eq.~(13) are large; e.g.,
for the halo mass $M_{\rm H}\approx 6\times 10^{12}\, M_\odot$ at $z\approx
2$ considered in Fig.~3 (corresponding to $\sigma_\star\approx 250$ km
s$^{-1}$), we have at the beginning of the declining phase a mass ratio
$M_{\rm res}/M_{\rm BH}\approx 3$, hence the viscous accretion rate amounts
to $\dot M_{\rm accr}\sim 10^2\, (\mathcal{R}_{\rm crit}/10^2)^{-1}\,
M_\odot$ yr$^{-1}$. The accretion rates required by the data on the FIR
detected fractions in X-ray selected AGN samples are lower than the simple
limits derived above, amounting to $\dot M_{\rm accr}\sim 5\,
(\epsilon/0.15)^{-1}\, M_\odot$ yr$^{-1}$. This may indicate that the fueling
mechanism is very complex, in that it does not depend solely on the amount of
gas still available after the QSO ejection, but also on various other
physical conditions. In fact, gas and dust spatial distribution, magnetic
fields, viscosity, cooling and heating, radiative pressure and additional
aspects have been recently introduced in hydrodynamical simulations to
capture the main features of the mass transfer toward the central BH (see
Narayan et al. 2012; Roth et al. 2012; Li et al. 2013; Liu et al. 2013). The
results we have derived from observations in the present paper can help
toward further, educated investigations.

\section{Summary and conclusions}

We have exploited the recent, wide samples of FIR selected galaxies
followed-up in X rays and of X-ray/optically selected AGNs followed-up in the
FIR band, along with the classic AGN and stellar luminosity functions at  
$z\gtrsim
1.5$, to infer the following scenario for the coevolution
of supermassive BHs and massive host galaxies ($M_{\rm H}\gtrsim 10^{12}\,
M_\odot$):

\begin{itemize}

\item the SFR in the host galaxy remains approximately constant for a time
    $\Delta t_{\rm burst}\sim 0.5-1$ Gyr and then declines abruptly because
    of the QSO feedback on a very short time scale $\tau_{\rm SFR}\approx
    \tau_{\rm ef}/3\approx \Delta t_{\rm burst}/40$;

\item part of the central ISM loses angular momentum and reaches the
    circumnuclear regions at a rate $\dot{M}_{\rm inflow}=\alpha_{\rm
    res}\times 10^{-3}\,\dot{M}_{\star}$, with typical values $\alpha_{\rm
    res}\approx 2$ and scatter $\sim 0.4$ dex reflecting those of the local
    BH to stellar mass ratio;

\item the early accretion onto a small seed BH, $M_{\rm seed}\sim
    10^{2-3}\,
    M_\odot$, occurs in a self-regulated regime with efficiency
    $\epsilon\approx 0.15$ and radiative power that can be just above the
    Eddington limit $\lambda\lesssim 4$, especially at the highest
    redshifts; thus the BH grows exponentially with an $e$-folding time
    $\tau\sim$ a few to several $10^7$ yr;

\item during the growth of the BH a massive reservoir forms, with a size of
    $20-100$ pc, close to the influence radius of the BH at the end of its
    evolution; the reservoir mass is as large as the final BH mass, ready
    to be delivered to the BH in the last $2-3$ $e$-folding times;

\item in the case of massive BHs the QSO feedback at its maximum exceeds
    the stellar feedback and is able to remove the ISM, stopping the star
    formation and the fueling of the reservoir; if the reservoir has enough
    gas a phase of supply-limited accretion follows, with luminosity (and
    Eddington ratio) exponentially declining over a timescale $\tau_{\rm
    AGN}\approx 2\, \tau_{\rm ef}$.

\end{itemize}

Interestingly, the time evolution of both the host galaxy and the AGN can be
characterized in terms of number of $e$-folding times; the galaxy/AGN
coevolution takes about $15$ $e$-folding times of the BH growth (see Figs.~1
and 3).

We have computed the FIR luminosity of the host galaxy and the bolometric
luminosity of the AGN/QSO as a function of the galactic age for different
formation redshifts and halo masses (see Figs.~1 and 3). We have exploited
them in association with the statistics of halo formation to reproduce:

\begin{itemize}

\item the luminosity function of FIR selected, high redshift galaxies (see
    Fig.~2), as well as the stellar mass function of their descendants,
    i.e., the passively evolving ETGs (see Fig.~4);

\item the luminosity function of X-ray and optically selected AGN/QSOs (see
    Figs.~11 and 18);

\item the available statistics on X-ray detections of FIR selected
    galaxies, and on FIR detections of X-ray/optically selected AGNs (see
    Figs.~5, 6, 7, 8, 10, 13, 14, 16);

\item the statistics of BH to stellar mass ratio in FIR selected galaxies
    and optically selected QSOs (see \S~4.1).

\end{itemize}

We stress that, in order to comply with the current FIR and X-ray data, any
model of coevolution must feature the time evolution of the SFR and BH
accretion rate outlined above, plus a relationship between SFR and halo mass
as a function of redshift in the line of Eq.~(4). We also stress the
importance of obtaining additional, accurate and well defined statistics on
the SFR in X-ray selected AGNs/QSOs and on the nuclear activity in FIR
selected galaxies, via coordinated observations from current and next
generation (sub-)mm and X-ray instruments. These
observations will be crucial to follow in greater detail the different stages
of the AGN/galaxy coevolution process, and in particular to pinpoint the
ratio between the duration timescale of the star formation and the
$e$-folding timescale of the BH accretion.

We have found that the ratio $L_{\rm FIR}/L_{\rm AGN}$ is the observational
parameter that characterizes the time evolution of the galaxy plus AGN
system (see Fig.~15, top panel); it is a decreasing function of the galactic
time and marks the evolution from the epoch when the luminosity budget is
dominated by the star formation to the epoch when the AGN/QSO takes over. The
FIR selection elicits objects with $L_{\rm FIR}/L_{\rm AGN}\gtrsim 0.3$ and
BH to stellar mass ratios $\Gamma$ smaller than the average local value
$\Gamma_0\approx 3\times 10^{-3}$. The primary X-ray selection picks out
objects spanning the whole range of the luminosity ratio; in particular, the
X-ray QSOs with large obscuration exhibit $0.3\lesssim L_{\rm FIR}/L_{\rm
AGN}\lesssim 5$. Instead, the optically selected QSOs show $0.01\lesssim
L_{\rm FIR}/L_{\rm AGN}\lesssim 0.3$, marking the decrease of the star
formation on a short timescales and the epoch of the reservoir exhaustion on
longer timescales; the bright objects feature a BH to stellar mass ratio
$\Gamma\gtrsim \Gamma_0$.

Our findings indicate the following scenario (see Fig.~15, bottom panel):
AGNs in massive halos begin their life as faint and obscured nuclei; then
they evolve into obscured QSOs visible mostly in the X-ray or in mid-IR
bands, and eventually become bright QSOs visible even in the optical
band. In this framework the role of the dust is crucial; it must form soon,
at the beginning of the host evolution, in order to yield the appropriate FIR
luminosity, and then it has to be largely removed at the beginning of the
optical bright AGN phase, in a couple of $e$-folding times. The nuclear
activity is detectable in hard X-ray (rest-frame) energies for longer times,
amounting to $4-5$ $e$-folding times; this is in line with the increasing
mean free path of the X-ray photons with increasing energy.

The time evolution of the star formation and BH accretion envisaged by our
model can be compared with other semi-analytical and numerical works appeared
in the recent literature. Khandai et al. (2012; see also Springel et al.
2005, Sijacki et al. 2009) exploited numerical simulations including galaxy
mergers and cold gas streams to investigate the growth of galaxies and hosted
supermassive BHs at high redshift. Their light-curves (see Fig.~2 of Khandai
et al. 2012) feature an early phase along which both the SFR and the BH
accretion rate rise, with the latter extremely rapid in attaining a few
$\times 10^{-1}\, M_\odot$ yr$^{-1}$; as a result, the AGN bolometric
luminosity exceeds $\sim 10^{45}$ erg s$^{-1}$ over a rather long period when
the SFR keeps sustained at more than $100\, M_\odot$ yr$^{-1}$, so as to
produce a very high fraction of X-ray detected AGNs in FIR selected hosts.
Subsequently, after the action of the AGN feedback, their light curves
feature a mild, slow decline in the SFR, with the BH accretion still at
appreciable values, so as to imply a high fraction of FIR detections in
X-ray/optically selected AGNs. Such high fractions are clearly at variance
with current observational data (see \S~3). Part of these problems can be
ascribed to: assumption of Bondi accretion strictly ceiled by the Eddington
limit, the ensuing requirements on large BH seed masses $\gtrsim 10^5\,
M_\odot$, and the recipe for mild AGN feedback on the large-scale star
formation activity.

Fanidakis et al. (2012; see also Cole et al. 2000, Baugh et al. 2005) exploit
a semi-analytic model that includes BH accretion by galaxy merging and disk
instabilities to compute the evolution of AGNs across cosmic times. Their
prescriptions (see Eqs.~1 and 2 of Fanidakis et al. 2012) imply a direct
proportionality between the BH accretion rate and the SFR. With their adopted
values for the fraction of stellar mass produced in the starburst that is
accreted onto the BH (their parameter $f_{\rm BH}\approx 5\times 10^{-3}$)
and for the ratio between the accretion and the bulge dynamical timescale
(their parameter $f_q\sim 1-10$), the resulting ratio of FIR to AGN
bolometric luminosity stays almost constant with time at values $L_{\rm
FIR}/L_{\rm AGN}\approx 2/f_q\sim 0.2-2$. This again implies a very high
X-ray detected fraction in FIR selected hosts. The problem can be mostly
ascribed to the absence of the reservoir in their model, and to a prominent
role of mergers in triggering the star formation and BH activity.

We have discussed the fueling of the reservoir (\S~5.2.1 and 5.2.2) and the
physics of BH accretion (see \S~5.2.3) in the light of numerical and analytic
models. We have shown that observations can be accounted for by models
predicting a self-regulated, super-Eddington accretion $\xi=\dot{M}_{\rm
accr}\, c^2/L_{\rm Edd}\approx 25$, associated to a radiative power only
slightly super-Eddington $\lambda=L/L_{\rm Edd}\lesssim 4$ and to a radiative
efficiency $\epsilon\lesssim 0.15$. However, this phase cannot be too short
to comply with the data, i.e., $\xi$ must not exceed $\sim 25$ by far. This
opens the theoretical issue of finding mechanisms able to provide accretion
rates exceeding the Eddington rate, but up to a maximum value. Subsequently,
the exponential decline of the luminosity required by the data implies the
Eddington ratio $\lambda$ to decrease. Correspondingly, following the
prescription of Eq.~(12) by Watarai (2000) and Li (2012) the radiative
efficiency $\epsilon=\lambda/\xi$ is expected to rapidly increase up to
$\epsilon\sim 0.3$ (see dotted line in Fig.~20). Numerical simulations honed
for this task can cast light on physical mechanisms at work to originate such
a behavior.

One of our major predictions is the formation of a reservoir in the
circumnuclear regions; the reservoir hosted by typical star forming galaxies
with $\dot{M}_{\star}\gtrsim 200\ M_{\odot}$ yr$^{-1}$ reaches a mass $M_{\rm
res}\gtrsim 3\times 10^8\, M_{\odot}$. In \S~5.1 we have pointed out that
these hosts are easily selected by (sub-)millimeter surveys such as
\textsl{Herschel}-ATLAS, {HerMES} and those performed with the \textsl{SPT}.
Many tens of these hosts are strongly lensed with amplification factors
$\gtrsim 10$ by intervening galaxies. In such cases, coordinated,
high-resolution observations in the (sub-)mm band with \textsl{ALMA} and in
the X-ray band with \textsl{Chandra} and the next generation of X-ray
telescopes will allow us to study in detail the evolution of the supermassive
BH and of its reservoir.

\begin{acknowledgements}
This work has been supported in part by the MIUR PRIN 2010/2011 `The dark
Universe and the cosmic evolution of baryons: from current surveys to
Euclid', by the INAF PRIN 2012/2013 `Looking into the dust-obscured phase of
galaxy formation through cosmic zoom lenses in the\textsl{Herschel}
Astrophysical Terahertz Large Area Survey', by ASI/INAF agreement No.
I/072/09/0, and by the INAF PRIN 2009 `New light on the early universe with
submillimeter spectroscopy'. We are grateful to the anonymous referee for
helpful comments and suggestions, and to A. Cavaliere, G.L. Granato, M.
Massardi, P. Salucci for interesting discussions. A.L. thanks SISSA for warm
hospitality.
\end{acknowledgements}

\begin{appendix}

\section*{Appendix. Computing detected fractions and luminosity functions}

In this Appendix we detail how we compute some quantities of interest in the
present work.

The fraction of objects $f_{\rm FIR\rightarrow X}$ selected in the FIR with
luminosity $L_{\rm FIR}$ within the bin [$L_{\rm FIR}^{\rm INF}$, $L_{\rm
FIR}^{\rm SUP}$] that are detected in X rays with an AGN luminosity $L_{\rm
X}$ above the limit $L_{\rm X}^{\rm LIM}$ is computed as:
\begin{equation}
f_{\rm FIR\rightarrow X}\equiv {\int{\rm d}M_{\rm H}~{\mathrm{d} N/\mathrm{d}
M_{\rm H}}\, \Delta t_{\rm FIR\rightarrow X}\over \int{\rm d}M_{\rm
H}~{\mathrm{d}N/\mathrm{d} M_{\rm H}}\, \Delta t_{\rm FIR}}~.
\end{equation}
Here ${\rm d}N/{\rm d}M_{\rm H} (M_{\rm H},z_{\rm form})$ is the halo mass
function, $\Delta t_{\rm FIR}[M_{\rm H},z_{\rm form}]$ is the time the object
spends at FIR luminosity $L_{\rm FIR}^{\rm INF}<L_{\rm FIR}[t|M_{\rm
H},z_{\rm form}]<L_{\rm FIR}^{\rm SUP}$ while $\Delta t_{\rm FIR\rightarrow
X}(M_{\rm H},z_{\rm form})$ is the time the object spends at FIR luminosity
$L_{\rm FIR}^{\rm INF}<L_{\rm FIR}[t|M_{\rm H},z_{\rm form}]<L_{\rm FIR}^{\rm
SUP}$ with X-ray luminosity ${L_{X}[t|M_{\rm H},z_{\rm form}]>L_{X}^{\rm
LIM}}$ above the detection limit. The average of a given quantity $Q[t|M_{\rm
H},z_{\rm form}]$ among the detected sources is computed as
\begin{equation}
<Q>_{\rm FIR\rightarrow X}\equiv {\int{\rm d}M_{\rm H}~{\mathrm{d}
N/\mathrm{d} M_{\rm H}}\, \int_{\Delta t_{\rm FIR\rightarrow X}}{\rm
d}t~Q(t)\over \int{\rm d}M_{\rm H}~{\mathrm{d}N/\mathrm{d} M_{\rm H}}\,
\Delta t_{\rm FIR}}~.
\end{equation}
The same formalism applies when computing the fraction of objects $f_{\rm
X\rightarrow FIR}$ selected in X rays and detected in the FIR, or the
fraction of objects $f_{\rm O\rightarrow FIR}$ selected in the optical band
(dominated by the QSO emission) and detected in the FIR.

In addition, the luminosity function in a given observational band is
computed as
\begin{equation}
\Phi(\log L,t) = \int^t{\rm d}t_{\rm form}\,\int{\rm d}\log M_{\rm H}\, {{\rm
d^2}N\over {\rm d}t_{\rm form}\,{\rm d}\log M_{\rm H}}\,
{{e^{-\log^2(L/L[t|M_{\rm H},t_{\rm form}])/2\,\sigma_{\log L}^2}}\over
\sqrt{2\pi\,\sigma_{\log L}^2}}~.
\end{equation}
In the above expression, ${\rm d^2}N/{\rm d}t_{\rm form}\,{\rm d}\log M_{\rm
H}$ is the halo formation rate as given by Lapi et al. (2013), $L[t|M_{\rm
H},t_{\rm form}]$ is the lightcurve from the model at observation time $t$
for a halo of mass $M_{\rm H}$ and formation time $t_{\rm form}$, and a
lognormal distribution of the luminosity around the average value $L[t|M_{\rm
H},t_{\rm form}]$ with scatter $\sigma_{\log L}$ has been assumed. For
further details we refer the reader to Lapi et al. (2006, 2011) and Cai et
al. (2013).

\end{appendix}

\clearpage
\begin{figure}
\epsscale{1}
\plotone{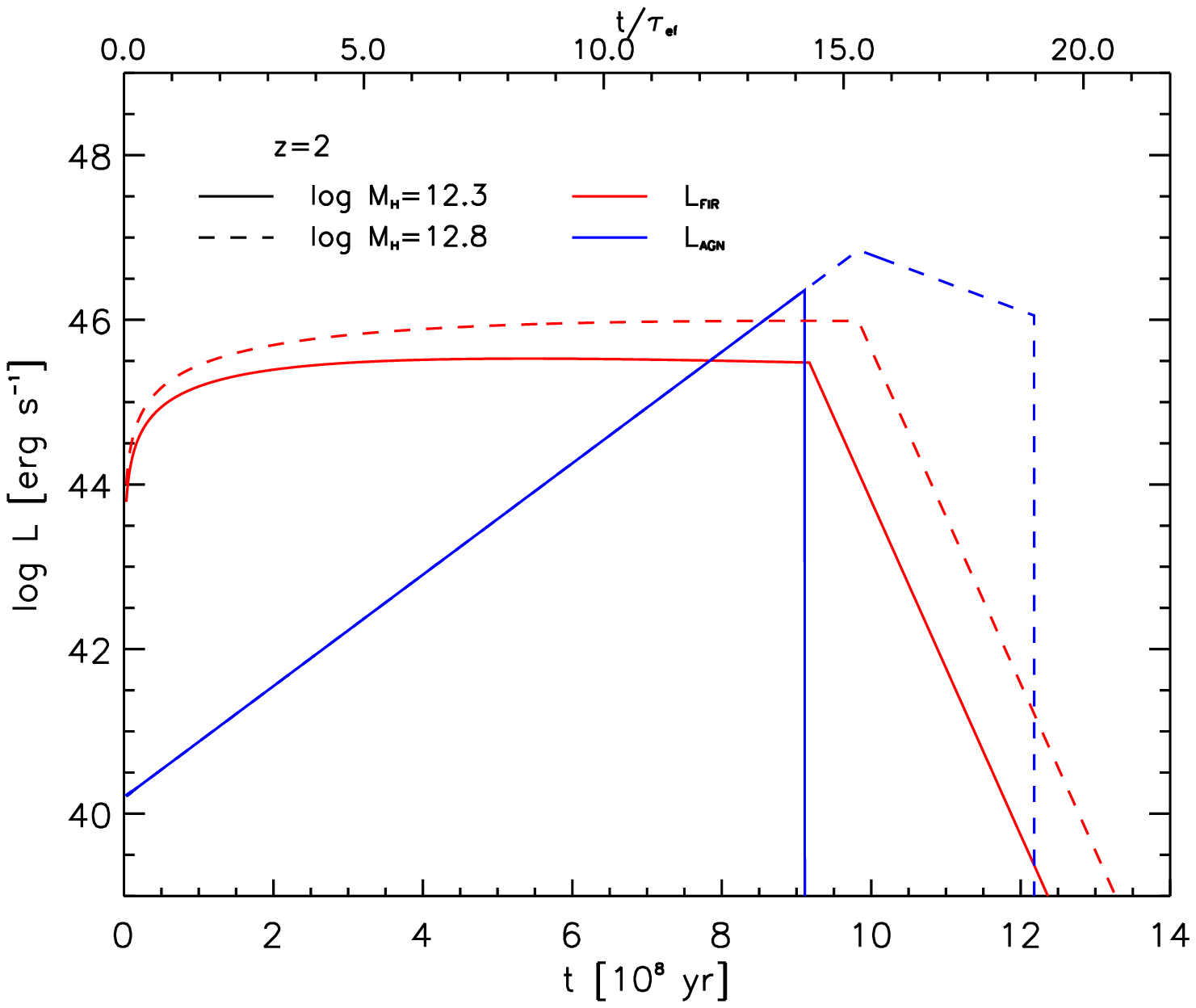}
\caption{Evolution of the bolometric AGN luminosity due to accretion (blue
lines) and of the FIR luminosity due to star formation (red lines) in
galaxies with halo mass $M_{\rm H}=2\times 10^{12}\,M_\odot$ (solid lines)
and $M_{\rm H}=6\times 10^{12}\,M_\odot$ (dashed lines) at redshift $z=2$.
The lightcurves are plotted as a function of the galactic age in units of
$10^8$ yr (lower scale) and of the $e$-folding time $\tau_{\rm ef}\approx
6\times 10^7$ yr (upper scale).}
\end{figure}

\clearpage
\begin{figure}
\epsscale{1}
\plotone{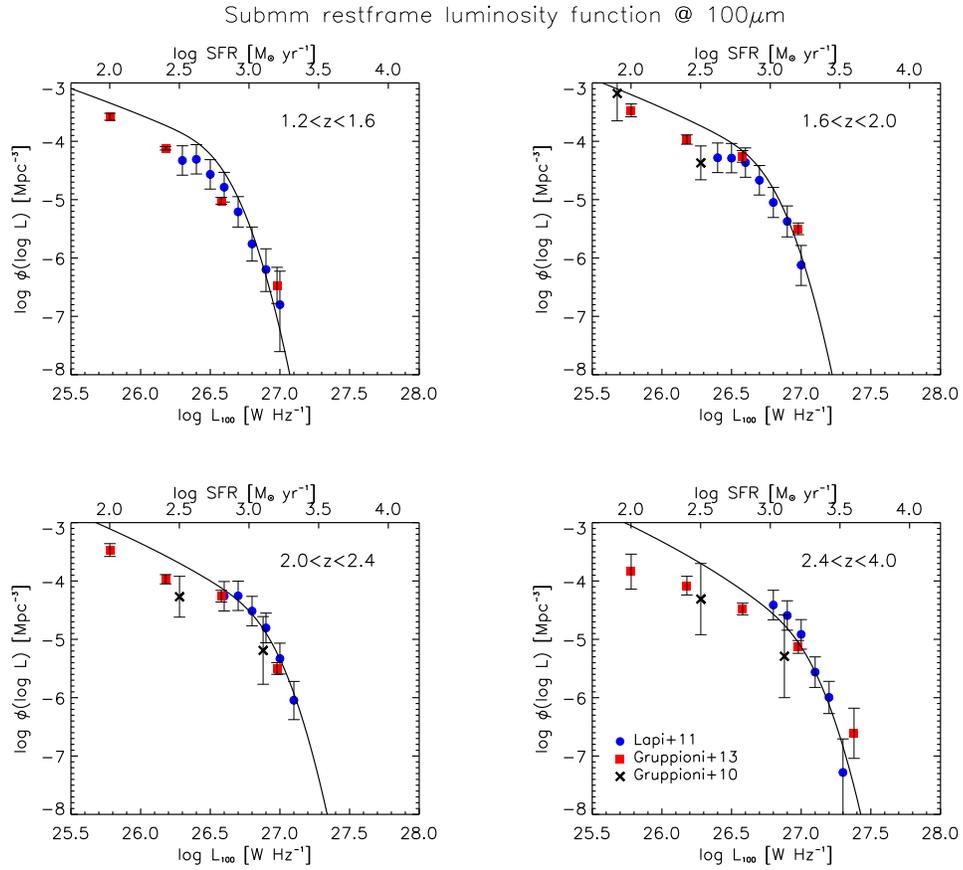}
\caption{FIR luminosity function at a restframe wavelength of $100\,\mu$m.
Each panel refers to a different redshift bin. The number density of sources
is plotted as a function of the luminosity at $100\, \mu$m restframe (lower
axis) and of the SFR (upper axis), having adopted a Chabrier IMF and the SED
of SMM J2135-0102, a typical high-redshift starbursting galaxy. The
predictions of the model (solid lines) are compared with the observational
data from the \textsl{Herschel}-ATLAS survey (Lapi et al. 2011) and the
\textsl{Herschel}-PEP survey (Gruppioni et al. 2010, 2013).}
\end{figure}

\clearpage
\begin{figure}
\epsscale{1}
\plotone{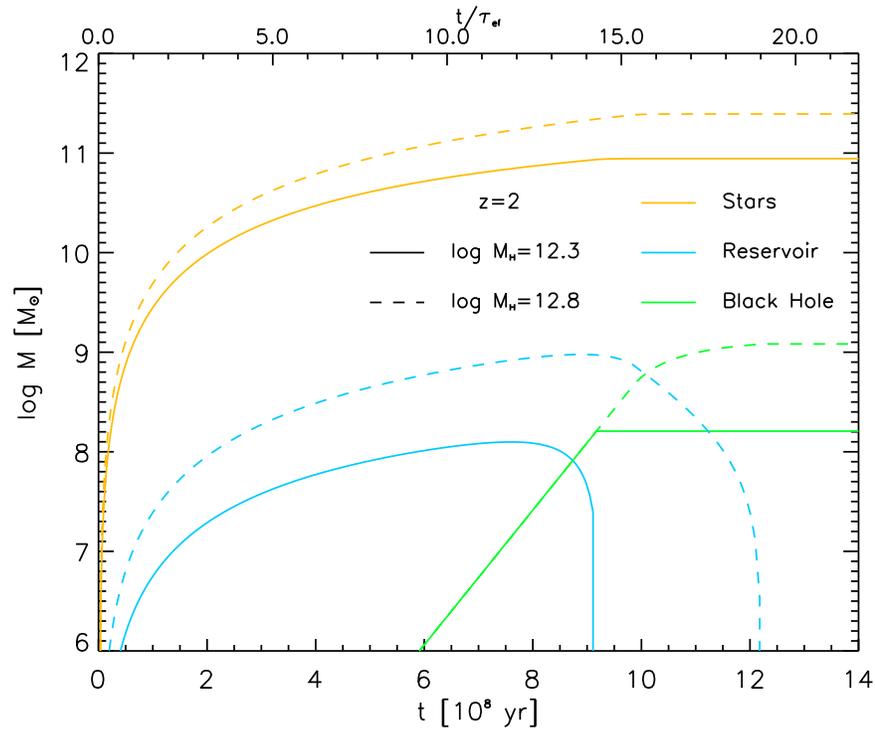}
\caption{Same as Fig.~1, but for the evolution of the stellar mass (orange
lines), reservoir mass (cyan lines), and BH mass (green lines).}
\end{figure}

\clearpage
\begin{figure}
\epsscale{1}
\plotone{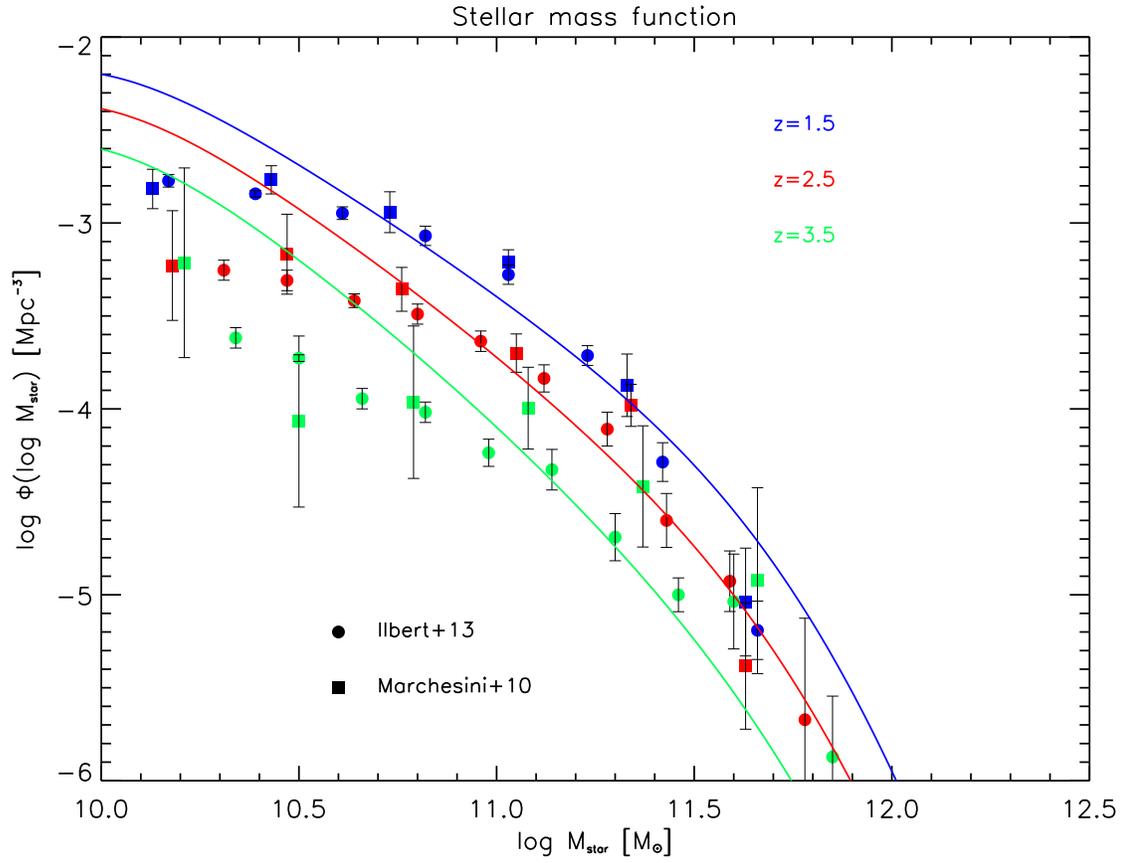}
\caption{Stellar mass function of passively evolving ETGs at redshifts
$z=1.5$ (blue), $2.5$ (red), and $3.5$ (green). Model predictions (solid
lines) are compared to the observational data from Marchesini et al. (2010,
squares) and Ilbert et al. (2013, circles).}
\end{figure}

\clearpage
\begin{figure}
\epsscale{1}
\plotone{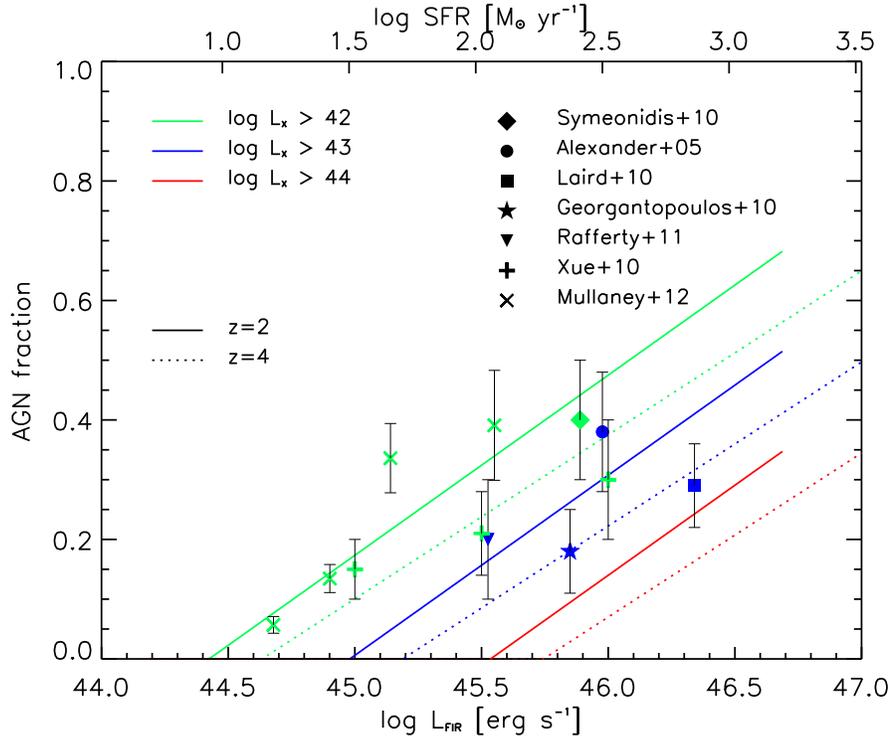}
\caption{Fraction of X-ray detected AGNs in FIR/$K$-band selected galaxies,
as a function of the FIR luminosity (lower scale) and of the SFR (upper
scale). Model predictions are provided for galaxies at $z=2$ (solid lines)
and $z=4$ (dotted lines) for different X-ray detection thresholds
$L_X=10^{42}$ (green), $10^{43}$ (blue), and $10^{44}$ erg s$^{-1}$ (red) in
the $2-10$ keV band; data (same color code) are from Alexander et al. (2005,
circles), Georgantopoulos et al. (2011, stars), Laird et al. (2010, squares),
Symeonidis et al. (2010; diamonds), Xue et al. (2010, plus signs), Rafferty
et al. (2011, triangles), Mullaney et al. (2012a, crosses).}
\end{figure}

\clearpage
\begin{figure}
\epsscale{1}
\plotone{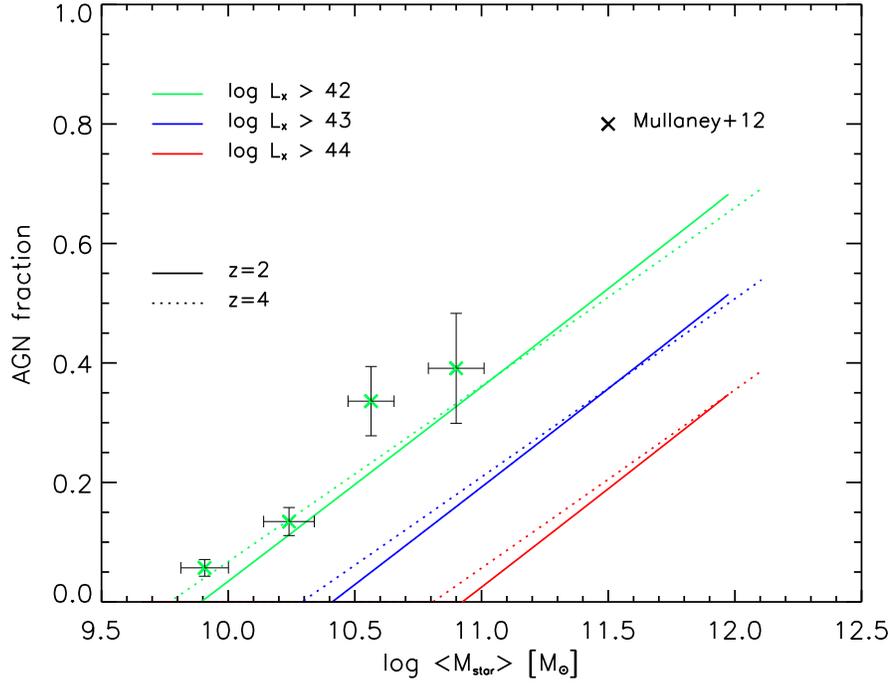}
\caption{Fraction of X-ray detected AGN in $K$-band selected galaxies, as a
function of the average stellar mass. Linestyles and color code as in Fig.~5.
Data are from Mullaney et al. (2012a, crosses).}
\end{figure}

\clearpage
\begin{figure}
\epsscale{1}
\plotone{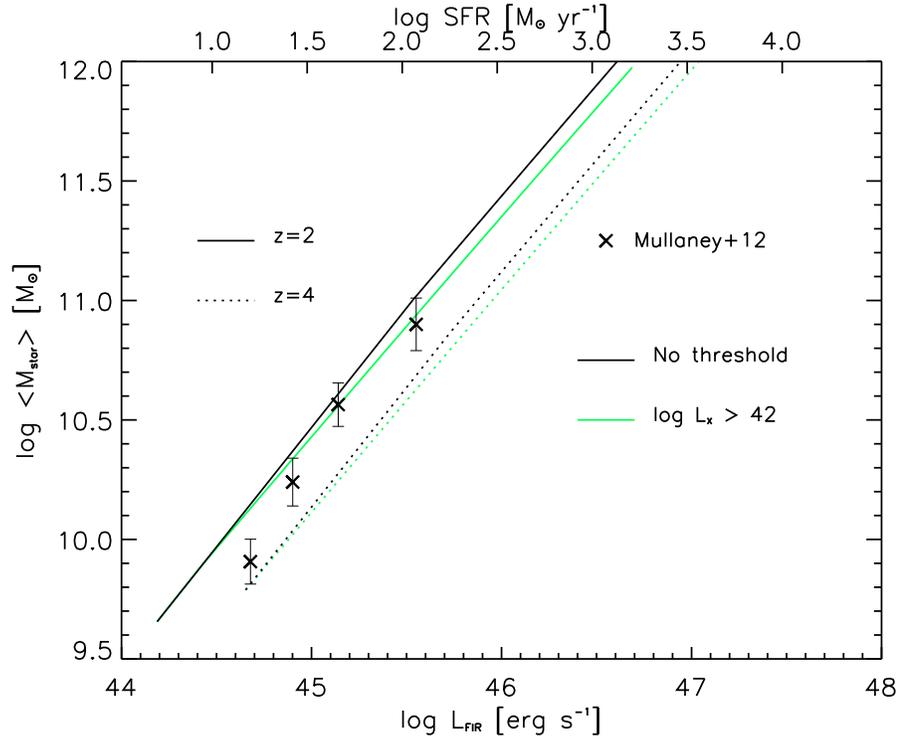}
\caption{Average stellar mass in FIR/$K$-band selected galaxies, as a
function of the FIR luminosity (lower scale) and of the SFR (upper scale).
Model predictions are provided at redshifts $z=2$ (solid lines) and $z=4$
(dotted lines), for an X-ray detection threshold $L_X=10^{42}$ erg s$^{-1}$
(green lines), and for no threshold (black lines); data (both for detected
and undetected sources, with the former dominating) are from Mullaney et al.
(2012a, crosses).}
\end{figure}

\clearpage
\begin{figure}
\epsscale{1}
\plotone{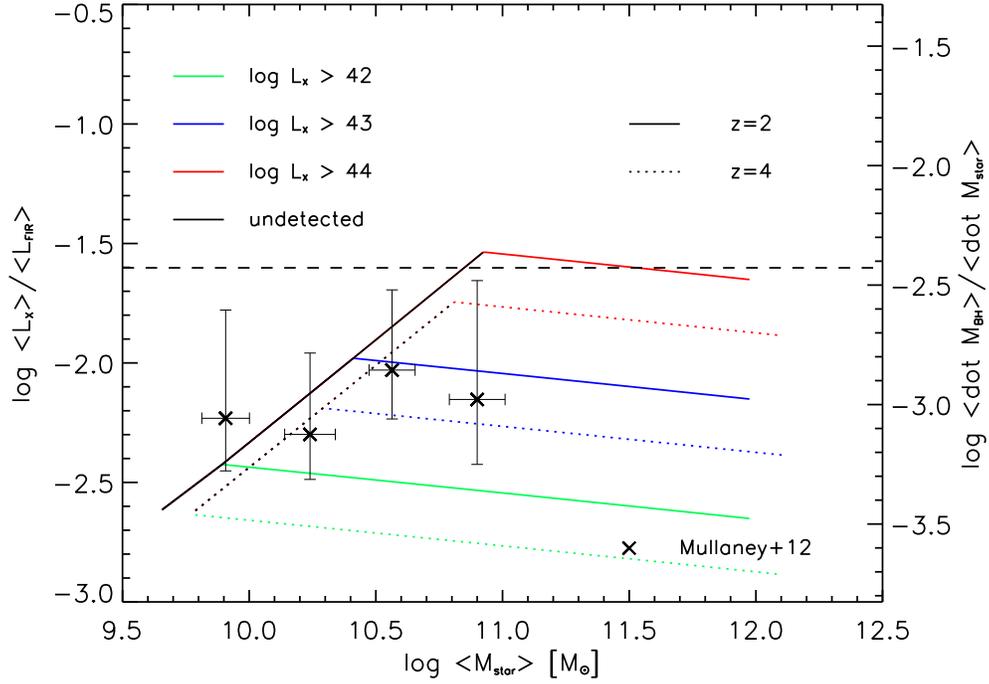}
\caption{Ratios of average X-ray to FIR luminosity (left scale) and of
average BH accretion rate to SFR (right scale) in $K$-band selected galaxies,
as a function of the average stellar mass. Linestyles and color code are as
in Fig.~5. The black lines show our predictions for undetected galaxies. The
dashed line corresponds to $L_{\rm FIR}=L_{\rm AGN}$ adopting $k_X=40$
(see Hopkins et al. 2007). Data are from Mullaney et al.
(2012a, crosses), and refer to a luminosity range a few $10^{42}\lesssim
L_X\lesssim 10^{43}$ erg s$^{-1}$.}
\end{figure}

\clearpage
\begin{figure}
\epsscale{1.0}
\plotone{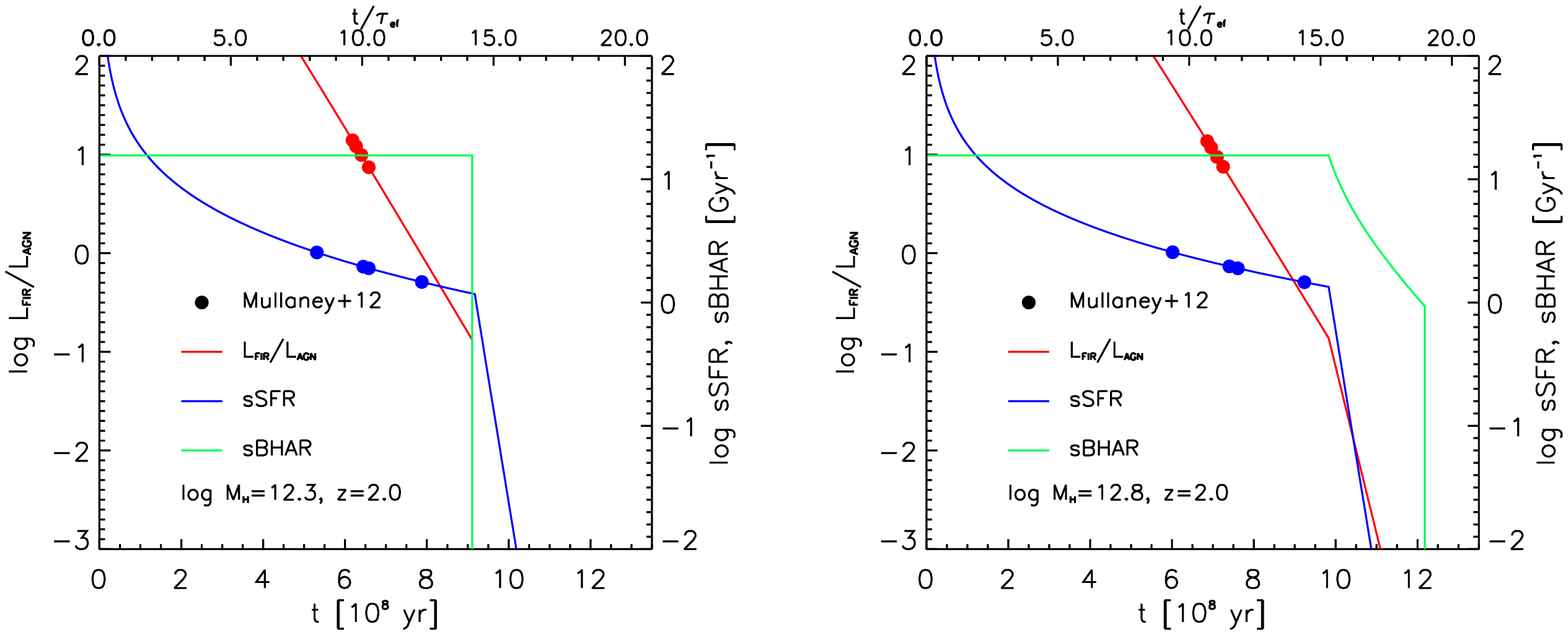}
\caption{Evolution of the bolometric FIR to AGN luminosity ratio (red lines,
left scale), of the specific SFR (blue lines, right scale), and of the
specific BH accretion rate (green line, right scale) in galaxies with halo
mass $M_{\rm H}=2\times 10^{12}\,M_\odot$ (left panel) and $M_{\rm H}=6\times
10^{12}\,M_\odot$ (right panel) at redshift $z=2$; the quantities are plotted
as a function of the galactic age in units of $10^8$ yr (lower scale) and of
the $e$-folding time $\tau_{\rm ef}\approx 6\times 10^7$ yr (upper scale).
The values of $L_{\rm FIR}/L_{\rm AGN}$ and of the sSFR observed by Mullaney
et al. (2012a, dots) are marked onto the respective model predictions.}
\end{figure}

\clearpage
\begin{figure}
\epsscale{1}
\plotone{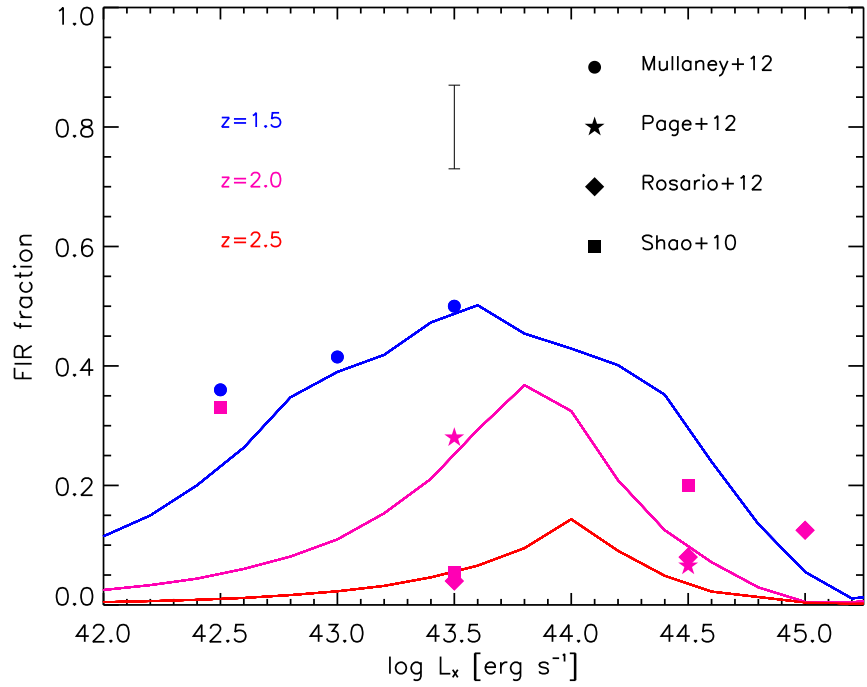} \caption{Fraction of FIR detected hosts in X-ray selected
AGNs, as a function of the X-ray luminosity. Model predictions (solid lines)
are provided at $z=1.5$ (blue), $z=2$ (magenta), and $z=2.5$ (red), for FIR
detection thresholds $\log L_{\rm FIR}=44.8$, $45.3$, and $45.8$ in erg
$s^{-1}$, respectively. Data (same color code) are from Shao et al. 2010
(squares), Mullaney et al. (2012b, circles), Page et al. (2012, stars), and
Rosario et al. (2012, diamonds); the typical error bar is also shown near the
legend.}
\end{figure}

\clearpage
\begin{figure}
\epsscale{1.1}
\center
\plotone{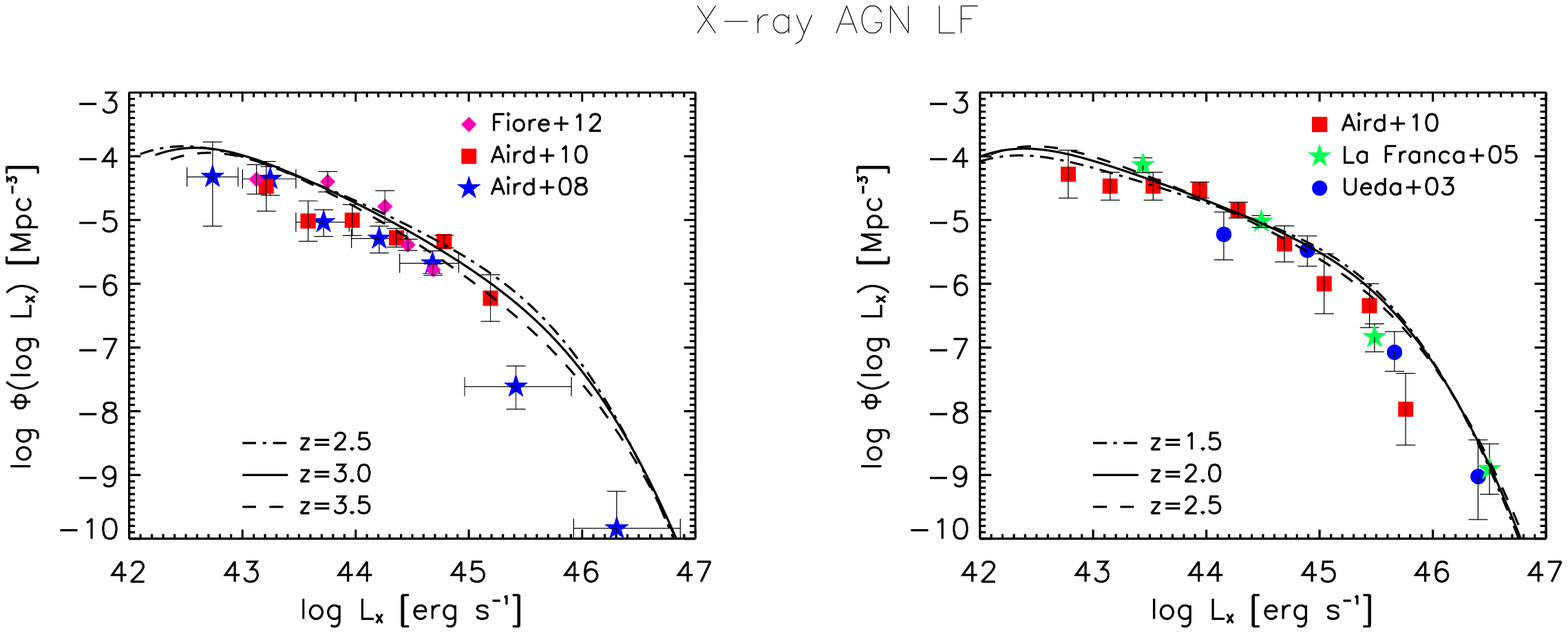}\caption{X-ray AGN luminosity function in two redshift bins
centered around $z=2$ (right panel) and $z=3$ (left panel). Model predictions
for various redshifts around these values (different linestyles) are compared
with the observations by Ueda et al. (2003, blue circles), La Franca et al.
(2005, green stars), Aird et al. (2008, blue stars), Aird et al. (2010, red
squares), Fiore et al. (2012a, magenta diamonds). The X-ray bolometric
correction by Hopkins et al. (2007) has been adopted.}
\end{figure}

\clearpage
\begin{figure}
\epsscale{1}
\plotone{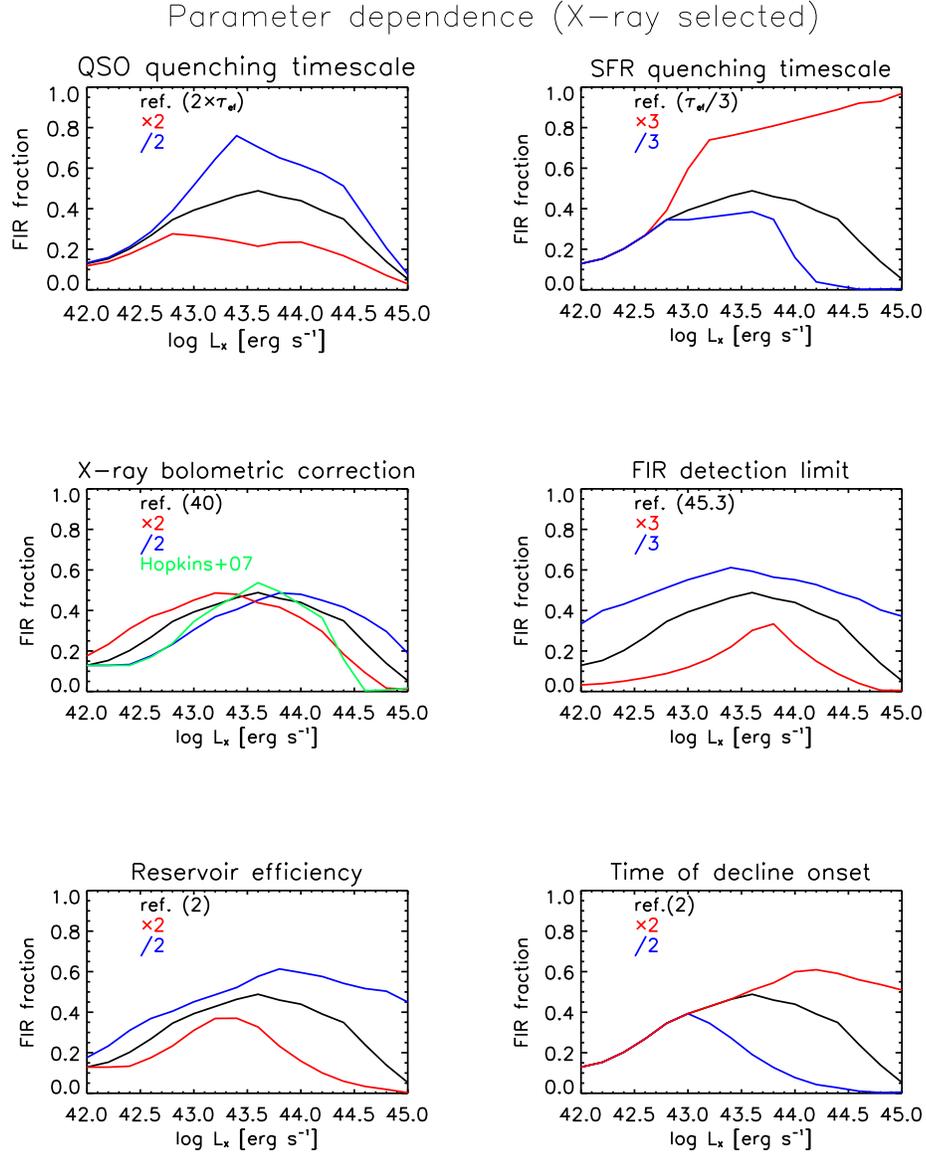}\caption{Dependence on various model parameters of the FIR
detected fraction of X-ray selected AGNs at $z=2$: quenching timescale of the
AGN emission $\tau_{\rm AGN}$ (top left); quenching timescale of the SFR
$\tau_{\rm
SFR}$ (top right); X-ray bolometric corrections $k_X$ (middle left); FIR
detection limit (middle right); reservoir efficiency $\alpha_{\rm res}$
(bottom left); parameter $n$ determining the onset of the declining phase
$t_{\rm QSO}+n\,\tau_{\rm ef}$ (bottom right; see text for details). In each
panel, three curves are shown: the black ones refer to a reference value of
the
parameter (reported in parenthesis in the legend), the red ones to an
increased
value, and the blue ones to a reduced value, as detailed in the figure. In
the
middle left panel, the additional green curve adopts the
luminosity-dependent bolometric correction by Hopkins et al. (2007).}
\end{figure}

\clearpage
\begin{figure}
\epsscale{1}
\plotone{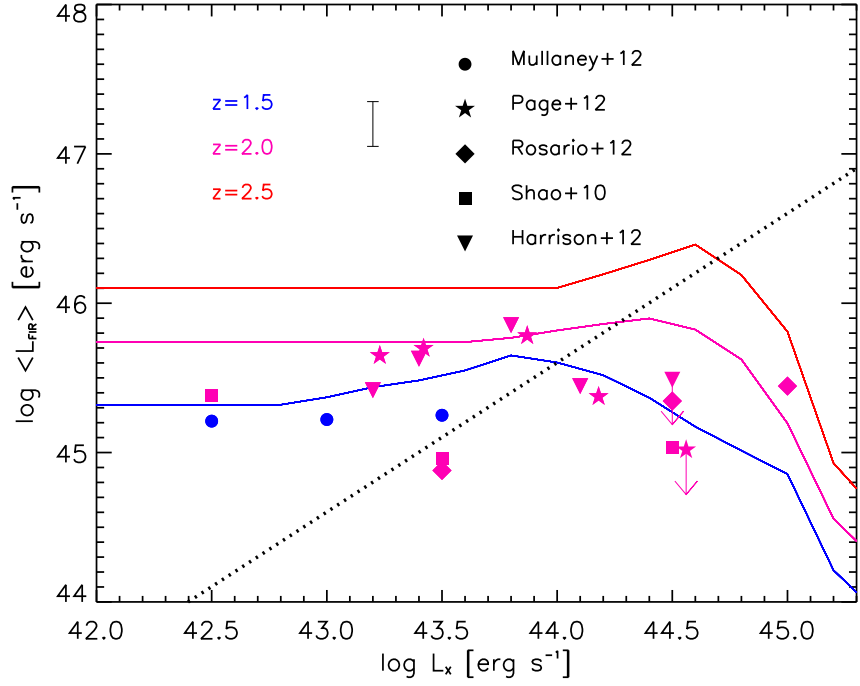} \caption{Average FIR luminosity of FIR detected hosts of
X-ray selected AGNs, as a function of the X-ray luminosity. Linestyles and
color code as in Fig.~10. Data are from Shao et al. (2010, squares), Harrison
et al. (2012, triangles), Mullaney et al. (2012b, circles), Page et al.
(2012, stars), and Rosario et al. (2012, diamonds). The dotted line
corresponds to $L_{\rm FIR}=L_{\rm AGN}$ where a
bolometric correction $k_X=40$ has been adopted.}
\end{figure}

\clearpage
\begin{figure}
\epsscale{1}\plotone{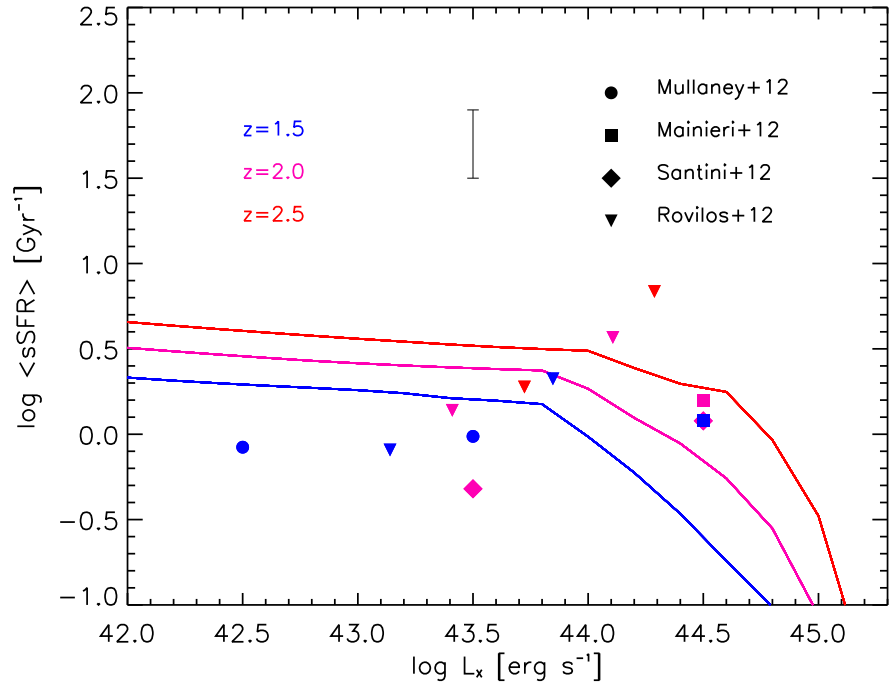} \caption{Average specific SFR of FIR
detected hosts of X-ray selected AGNs, as a function of the X-ray luminosity.
Linestyles and color-code as in Fig.~10. Data are from Mainieri et al. (2012,
squares), Mullaney et al. (2012b, circles), Rovilos et al. (2012, triangles),
and Santini et al. (2012, diamonds).}
\end{figure}

\clearpage
\begin{figure}
\epsscale{0.7}
\plotone{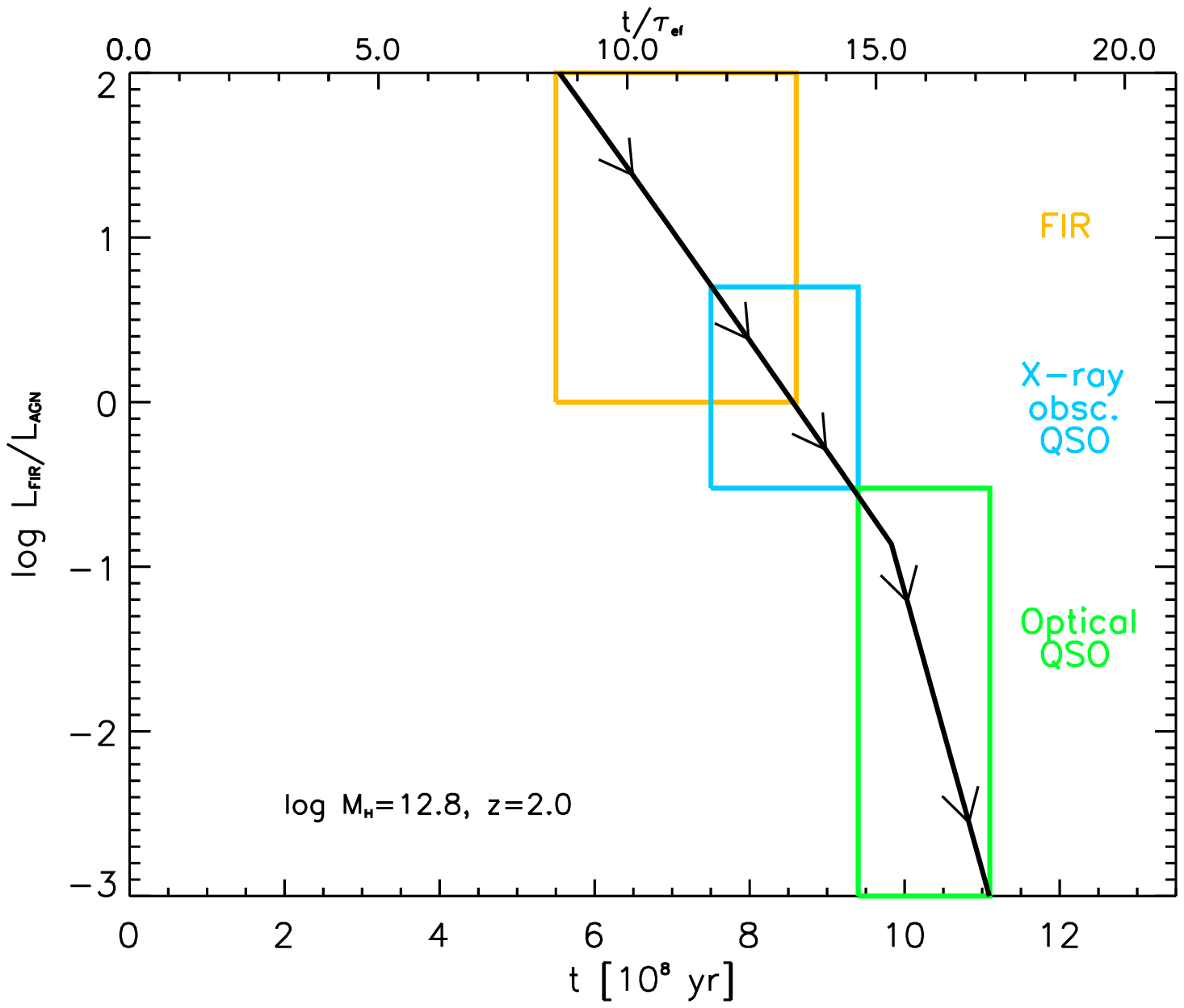}\\\plotone{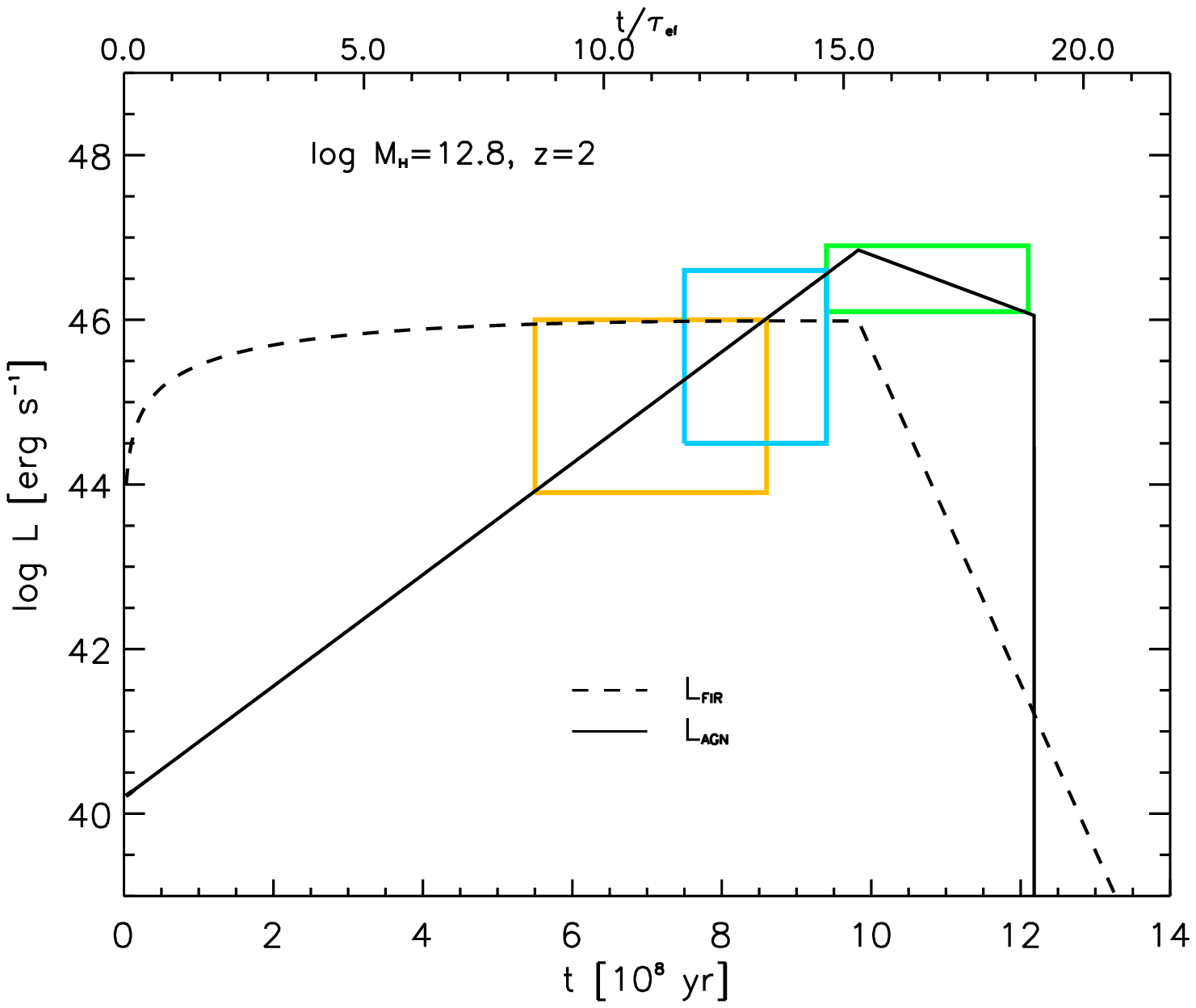} \caption{The top panel shows a
schematic illustration of the different phases, marked by the colored boxes,
in the galaxy/AGN coevolution envisaged by the model, as traced by the ratio
$L_{\rm FIR}/L_{\rm AGN}$. Specifically, the orange box refers to the FIR
bright phase, the cyan box to the obscured X-ray QSO phase, and the green box
to the optically bright QSO phase. Note also that an X-ray selection unbiased
to obscuration can pick up objects with very different values of $L_{\rm
FIR}/L_{\rm AGN}$, cf. Fig.~13. The bottom panel shows the placement of the
same
phases on the FIR and AGN lightcurves.}
\end{figure}

\clearpage
\begin{figure}
\epsscale{1}
\plotone{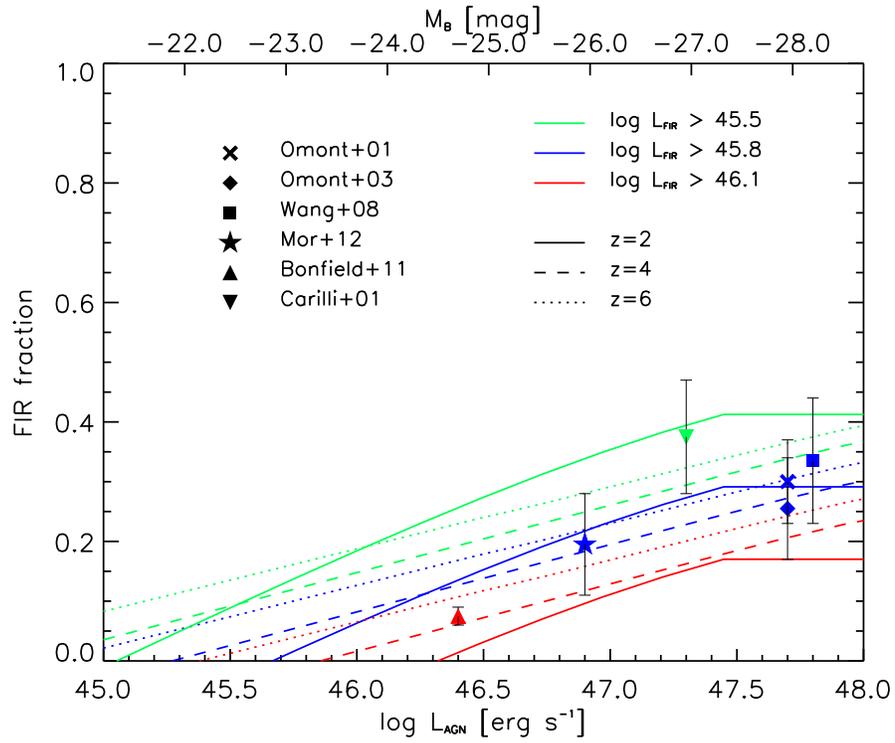}\caption{Fraction of FIR detected hosts of optically
selected AGNs, as a function of the AGN bolometric luminosity (lower scale)
and of the $B$-band magnitude (upper scale). Model predictions are provided
at redshift $z=2$ (solid lines), $4$ (dashed lines), and $6$ (dotted lines),
for different values of the FIR detection threshold $\log L_{\rm FIR}=45.5$
(green lines), $45.8$ (blue lines), and $46.1$ (red lines) in units of erg
s$^{-1}$. Data (same color code) are from Carilli et al. (2001, reversed
triangle), Omont et al. (2001, cross), Omont et al. (2003, diamond), Wang et
al. (2008, square), Bonfield et al. (2011, triangle), and Mor et al. (2012,
star).}
\end{figure}

\clearpage
\begin{figure}
\epsscale{1}
\plotone{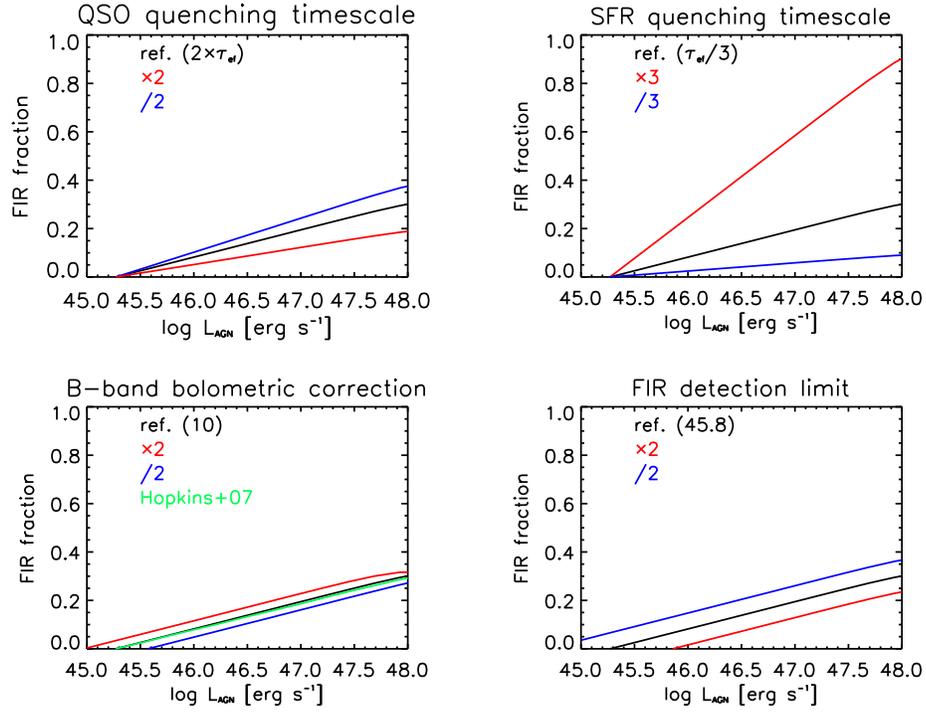} \caption{Dependence on various model parameters of the FIR
detected fraction of optically selected AGNs at $z=4$: quenching timescale of
the AGN $\tau_{\rm AGN}$ (top left); quenching timescale of the SFR
$\tau_{\rm SFR}$ (top right); X-ray bolometric corrections $k_X$ (bottom
left); FIR detection threshold (bottom right). In each panel, three curves
are shown: the black ones refer to the fiducial value of the parameter
(reported in parenthesis), the red ones to an increased value, and the blue
ones to a reduced value, as detailed in the legend. In the bottom left panel,
the additional green curve is for the luminosity-dependent bolometric
correction by Hopkins et al. (2007).}
\end{figure}

\clearpage
\begin{figure}
\epsscale{1.1}
\center
\plotone{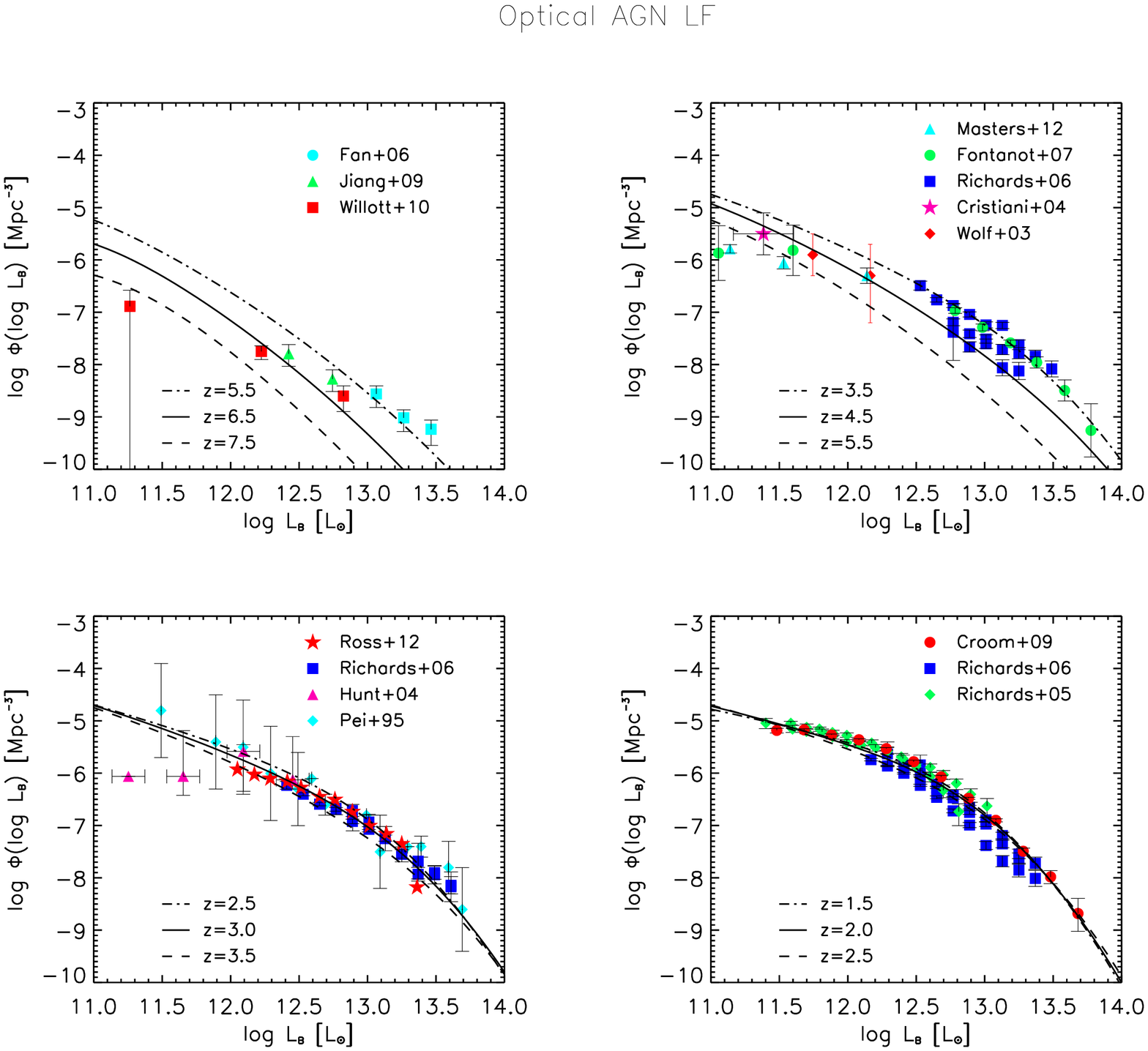}
\caption{Optical QSO luminosity function in four redshift bins centered
around $z=6.5$ (top left), $4.5$ (top right), $3$ (bottom left), and $2$
(bottom right). Model predictions for various redshifts around these values
(different linestyles) are compared with the observations by Pei (1995, cyan
diamonds), Wolf et al. (2003, red diamonds), Cristiani (2004, magenta star),
Hunt et al. (2004, magenta triangles), Richards et al. (2005, green
diamonds), Fan et al. (2006, cyan circles), Richards et al. (2006, blue
squares), Fontanot et al. (2007, green circles), Croom et al. (2009, red
circles), Jiang et al. (2009, green triangles), Willott et al. (2010, red
squares), Masters et al. (2012, cyan triangles), Ross et al. (2013, red
stars). The optical bolometric correction by Hopkins et al. (2007) has been
adopted.}
\end{figure}

\clearpage
\begin{figure}
\epsscale{1}
\plotone{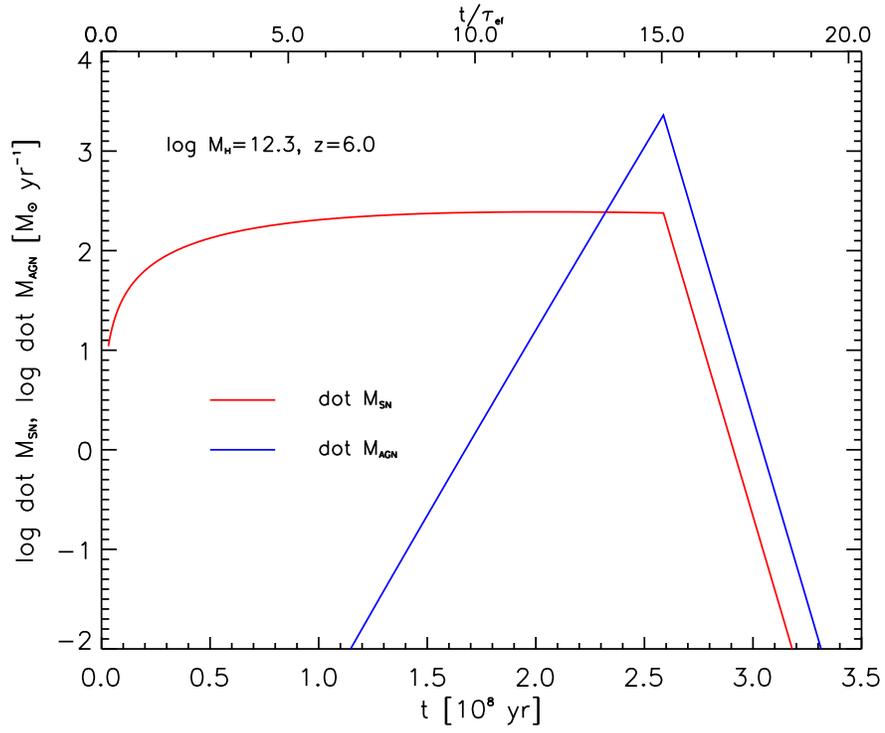} \caption{Evolution of the cold gas outflow rate due to SN
(red line) and AGN feedback (blue lines) for a galaxy with halo mass $M_{\rm
H}=2\times 10^{12}\,M_\odot$ at redshift $z=6$; the curves are plotted as a
function of the galactic age in units of $10^8$ yr (lower scale) and of the
$e$-folding time $\tau_{\rm ef}=2\times 10^7$ yr (upper scale).}
\end{figure}

\clearpage
\begin{figure}
\epsscale{1}
\plotone{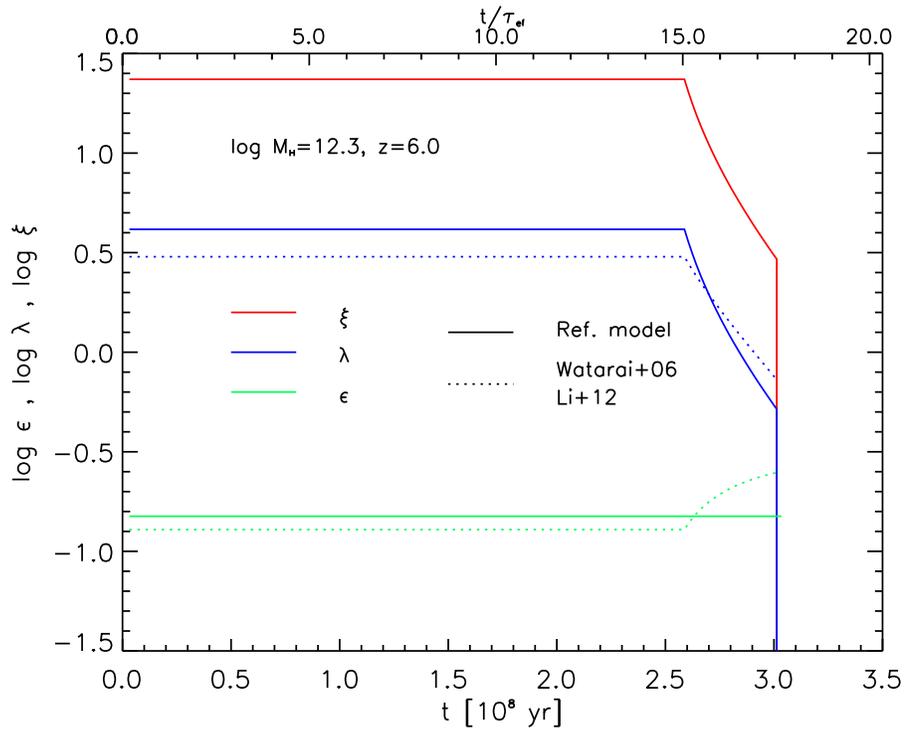} \caption{Evolution of the ratio between the BH mass
accretion to the Eddington rate $\xi=\dot M_{\rm accr}\, c^2/L_{\rm Edd}$
(red line), of the Eddington ratio $\lambda=L/L_{\rm Edd}$ (blue line), and
of the radiative efficiency $\epsilon=\lambda/\xi$ (green line) in a galaxy
with halo mass $M_{\rm H}=2\times 10^{12}\,M_\odot$ at redshift $z=6$; the
evolution curves are plotted as a function of the galactic age in units of
$10^8$ yr (lower scale) or of the $e$-folding time $\tau_{\rm ef}=2\times
10^7$ yr (upper scale). The solid lines refer to the model while
the dotted lines to the prescriptions by Watarai (2006) and Li
(2012).}
\end{figure}

\end{document}